\documentclass[final,1p,times]{elsarticle}

\usepackage{amssymb, amsmath}
\usepackage{graphicx,epsf}
\usepackage{color}

\journal{Nuclear Physics B}

\begin{document}

\begin{frontmatter}

%\title{Large conformal Goldstone mode fluctuations in the SYK-model}

\title{Quantum ergodicity in the SYK model} 

\author[label1]{Alexander Altland}
\author[label1]{Dmitry Bagrets}
\address[label1]{Institut f\"ur Theoretische Physik, Universit\"at zu K\"oln,
Z\"ulpicher Stra\ss e 77, 50937 K\"oln, Germany}

%\author{}
%\address{}

\begin{abstract}
We present a replica path integral approach describing the quantum chaotic dynamics
of the SYK model at large time scales. The theory leads to the identification of non-ergodic collective modes which relax and eventually give way  to an
ergodic long time regime (describable by random matrix theory). These
modes, which play a role conceptually similar to the diffusion modes of dirty metals,
carry quantum numbers which we identify as the generators of the Clifford algebra:
each of the $2^N$ different products that can be formed from $N$ Majorana operators
defines one effective  mode. The competition between a decay rate quickly growing in
the order of the product and a density of modes exponentially growing in the same
parameter explains the characteristics of the system's approach to the ergodic
long time regime. We probe this dynamics through various spectral correlation
functions and obtain favorable agreement with existing numerical data.
\end{abstract}


\end{frontmatter}


\section{Introduction}
\label{sec:Intro}
The Sachdev-Ye-Kitaev (SYK) model \cite{Sachdev:1993,Kitaev:2015} has become a
paradigm of hard quantum chaos in strongly interacting quantum matter. Standing in
the tradition of a general class of random interaction
models~\cite{French:1970,French:1971,Bohigas:1971,Bohigas:1971a}, it is a system of $N$
(Majorana) fermions, $\chi_i$ with $i=1,\ldots,N$, subject to a four-fermion
interaction
\begin{align}
    \label{eq:Hamiltonian}
    H=\sum_{ijkl}^N J_{ijkl}\,\chi_i\chi_j\chi_k\chi_l,
\end{align}
with Gaussian distributed random matrix elements $J_{ijkl}$ of zero mean and a
variance given by  $\langle |J_{ijkl}|^2\rangle=6J^2/N^{3}$. The model is
known~\cite{Kitaev:2015} to show hard many body quantum chaos at all time scales. For
`semiclassically short' times, chaos manifests itself in exponentially decaying
correlations, as described by out of time correlation functions~\cite{Larkin:1969,
Aleiner:2016, Maldacena2:2016, Maldacena1:2016, Bagrets:2017, Kitaev:2017}. In the
complementary regime of ultra-long times, of the order of the inverse of the many
body level spacing, chaos is diagnosed via quantum level repulsion otherwise found
for random matrix theory (RMT) ensembles~\cite{You:2016}. However, a question that
has not really been answered so far is at what time or energy scales the system
actually becomes \emph{ergodic}. Relatedly, the nature of the system's effectively
irreversible dynamics prior to entering the asymptotic ergodic long time regimes
remains unclear. To motivate the question on a simpler example, the dynamics of a
diffusive $d$-dimensional metal of linear extension $L$ is chaotic at all time scales
(exceeding the elastic scattering time.) However a crossover to \emph{ergodic} long
time dynamics takes place only at times $t>t_{\mathrm erg}=L^2/{\cal D}$ exceeding the
classical diffusion time through the system, where ${\cal D}$ is the diffusion constant.
That time scale is called the ergodic time, or, in the specific context of dirty
metals, the Thouless time. At time scales shorter than $t_{\mathrm erg}$ the dynamics
of the system is governed by  diffusion modes relaxing in time. Technically, these are 
eigenmodes of the diffusion operator, and they  are
labeled by a set of (`momentum') quantum numbers $q=n 2\pi/L$, where
$n=(n_1,\dots,n_d)$ is a vector of integers, and ${\cal D} |q|^2$ defines the decay rates.
The inverse of the lowest non-vanishing of these scales, ${\cal D} (2\pi/L)^2$ defines the
Thouless energy.

In this paper, we address analogous questions for the SYK model: what is its ergodic
time, and what is the nature of the relaxation modes prevailing at shorter scale? Can
these modes be classified by effective `quantum numbers', and if so, what is the
density of these modes? Finally, what are the observable consequences in spectral
correlation functions? We will provide answers to these questions and test their
validity by  comparison to existing numerical data. Specifically, there are two
numerical analyses providing  test criteria for our approach: in
Ref.~\cite{Garcia:2016} the \emph{spectral number variance}, $\Sigma_2(\epsilon)$,
i.e. the statistical variation in the number of many body levels contained in an
energy window of width $E$ has been obtained for systems of fermion number up to
$N=34$. For energies beyond an $N$-dependent time scale (which was difficult to
estimate quantitatively on the basis of the available data but conjectured to be an
algebraic power of the band-width) deviations from the results of random matrix
ensemble number variances were seen. These deviations signal the breakdown of
ergodicity and their quantitative computation is one of the objectives of our
analysis. In Ref.~\cite{Cotler:2017}  the \emph{spectral form factor}, $K(\tau)$,
i.e. the Fourier transform of the energy  dependent spectral two-point correlation
function, $R_2(\epsilon)$ (for the concrete definition of these functions, see the
next section), was computed for systems of different size. While the long time
profile showed a ramp structure characteristic for RMT ensembles, universal
deviations were observed for shorter times
(see also~\cite{Li:2017,Hunter_Jones:2017,Cotler:2017a,delCampo:2017} 
for related  studies). The quantitative analytic reproduction of
the non-ergodic contributions to the number variance and the form factor, and the
demonstration that they originate in the same set of relaxation modes sets a
stringent test for the validity of our analysis.

In this paper, we will approach the SYK model from a perspective different from that
of previous analyses. The idea is to consider its Hamiltonian as a random \emph{first
quantized} operator (a random matrix) acting in the $2^{N/2}$-dimensional Hilbert
space of the system. This matrix is \emph{sparse} in that it contains only
algebraically many independent matrix elements, compared to a rank increasing
exponentially in $N$. The comparatively low entropy contained in this structure is
responsible for the phenomenological deviations from   maximum entropy random matrix
Hamiltonians defined through a full set of i.i.d. distributed matrix elements.
Methodologically, the advantage gained from the  first quantized perspective is that
powerful field theoretical methods developed for random single particle problems
become applicable to the present system. Conceptually, this approach provides insight
into the question how many-body quantum chaos seeded into a large Hilbert-Fock space
via the `few' interaction matrix elements works its way through an exponentially
large phase volume to eventually stabilize an ergodic phase.

We will start  in the next section with a brief review of spectral correlations in
random quantum single particle systems. In view of numerous analogies this will be
instructive  and introduce the appropriate language for our later discussion of the
SYK problem. In the second part of
section~\ref{sec:qualitative_discussion_and_summary_of_results} we summarize our main
results and compare to earlier numerical studies. In
section~\ref{sec:replica_field_theory} we introduce the field theoretical framework
for the quantitative analysis and in section~\ref{sec:StationaryPhaseAnalysis}
formulate a mean field analysis. In sections~\ref{sec:FluctuationsRMT}
and~\ref{sec:FluctuationsMassive} we discuss the types of fluctuations relevant for
the description of the ergodic sector and the relaxation modes, respectively.
Section~\ref{sec:FluctuationsMassiveNonlinear} contains the technically most involved
part of the program,  the demonstration of the absence of non-linear corrections to
the mean field results. (In view of the high dimensionality of the present problem
and a correspondingly large `phase volume' of fluctuations, this step is essential.
However, readers primarily interested in results may skip this part.) We conclude in
section~\ref{sec:summary_and_discussion}. Technical parts of the discussion are
relegated to several appendices.

\begin{figure}[t]
\centering{\includegraphics[width=14.0cm]{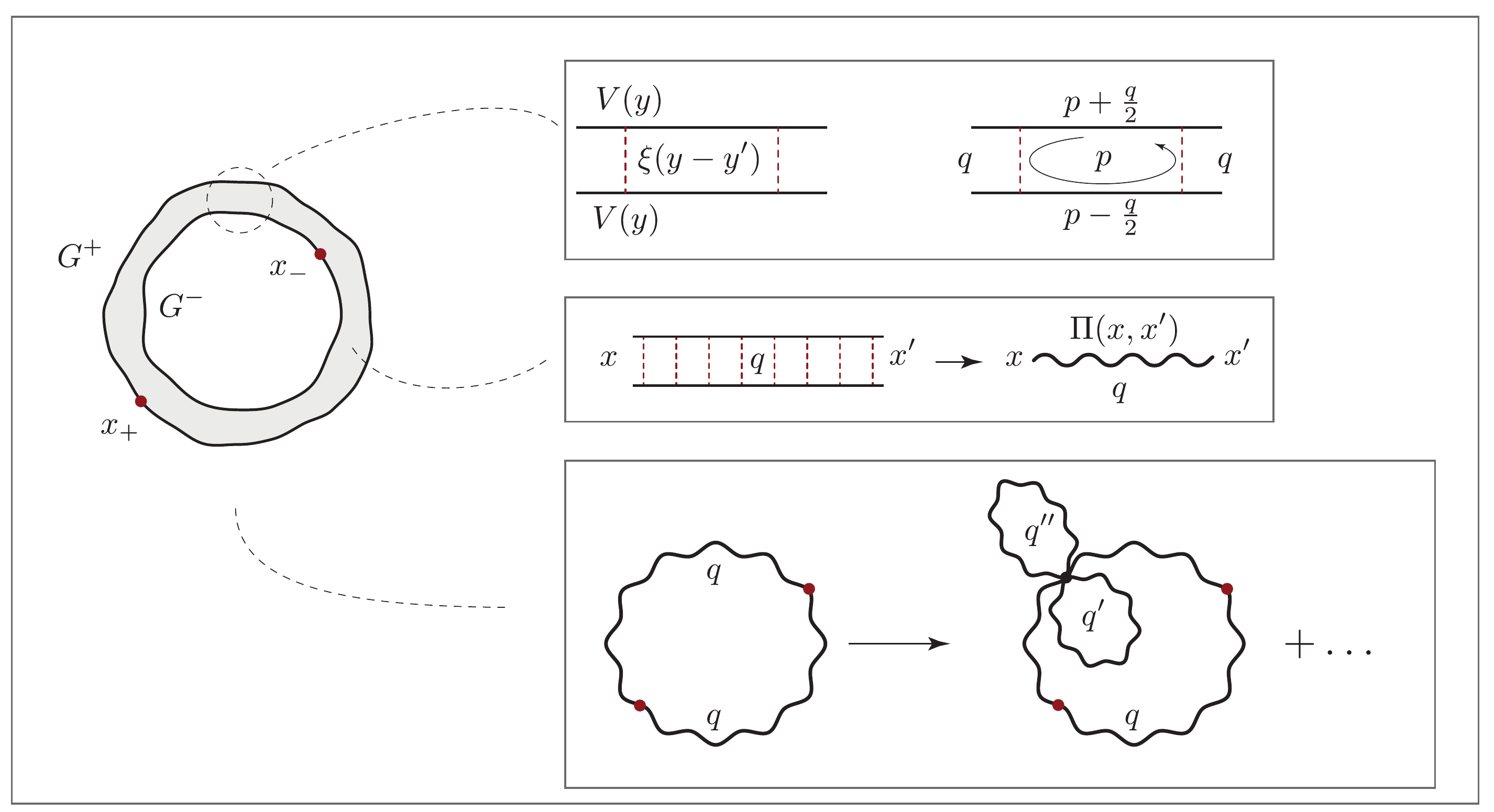}}
\caption{Left: Semiclassical representation of the spectral two-point function
through Green function amplitudes. Inset top: microscopic structure of scattering
vertex in coordinate (left) and momentum (right) representation. Inset center:
abbreviated representation of momentum conserving two particle mode. Inset bottom:
spectral correlation function as one loop diagram involving two modes. At low
energies higher order loop processes gain importance. Further discussion, see text.}
\label{fig:SingleParticleSpectralStatistics} 
\end{figure}

\section{Qualitative discussion and summary of results} % (fold)
\label{sec:qualitative_discussion_and_summary_of_results}

Consider a stochastic quantum system described by a statistical ensemble of
Hamiltonians $H$. Its spectral fluctuations at a characteristic energy $E$ are
described by the two point function $R_2(\omega)\equiv \Delta^2 \left\langle
\rho(E+\frac{\omega}{2})\rho(E-\frac{\omega}{2})\right\rangle_c$, where $\Delta\equiv \Delta(E)\equiv
\langle \rho(E)\rangle^{-1}$ is the average level spacing at $E$, $\rho$ the density
of states (DoS) and $\left\langle \dots\right\rangle_c$ is a cumulative (`connected')
average over randomness. Prominent quantities derived from the two-point function
include the Fourier transform, or \emph{spectral form factor} $K(\tau)\equiv
\frac{1}{\Delta}\int d\omega\, R_2(\omega)e^{-i\frac{2\pi\omega}{\Delta} \tau}$  and
the \emph{number variance}, $\Sigma_2(\epsilon)\equiv \langle
N(\epsilon)^2\rangle_c$, where $N(\epsilon)=\int_{E-\epsilon/2}^{E+\epsilon/2}d
\epsilon'\rho(\epsilon')$ is the number of levels contained in a strip of width
$\epsilon$.

In view of the relation $\rho(\epsilon)=-\frac{1}{\pi}{\rm Im}\,\textrm{tr}(G^+(\epsilon))$, where $G^\pm(\epsilon)=(\epsilon^\pm -H)^{-1}$ is the resolvent, the full information on all these quantities is contained in the two-point function
\begin{align}
    \label{eq:CDef}
    C(\omega)\equiv \langle \textrm{tr}(G^+(\epsilon+ \tfrac{\omega}{2}))\textrm{tr}(G^{-}(\epsilon - \tfrac{\omega}{2})) \rangle_c,
\end{align}
where the (generally weak) dependence of $C$ on the center energy $\epsilon$ is
suppressed and we noted that connected averages between Green functions of the same
causality vanish, $\left\langle G^{\pm}G^{\pm}\right\rangle_c=0$. A semiclassical
cartoon of the situation is shown in Fig.~\ref{fig:SingleParticleSpectralStatistics},
where the black lines are Green function propagators, and the dots represent the
common starting and end state, $x_\pm$, in $\textrm{tr}(G^\pm)=\sum_{x_\pm}
G^\pm(x_\pm,x_\pm)$. The ring shaped structure indicates that the two propagators
must remain piecewise close to each other to remain statistically correlated. The
top inset illustrates the situation for the case of a Gaussian potential
$\left\langle V(y)V(y')\right\rangle\sim
\xi(y-y')$ with finite range correlation function $\xi$. The scattering processes off
fluctuations $V(y)$ are indicated by dashed lines, and a line connecting two
propagator amplitudes represents the average over a product of two of these.  In this
case, the extent of $\xi$ sets the  tolerance for deviations between the Feynman
amplitudes. The effective two-particle mode emerging in this way (inset middle)
defines a quantum stochastic process which, after averaging over the randomness, is
governed by an effective master equation $(\partial_t
-\mathcal{O})\Pi(x,x',t)=\delta(x,x')\delta(t)$. Here, $\Pi(x,x',t)$ is the
probability of pair propagation between two points $x$ and $x'$ in time $t$ and
$\mathcal{O}$ a local operator whose specifics depend on the context. For example, in
the particular case of an extended medium with Gaussian randomness,
$\mathcal{O}= {\cal D} \partial_x^2$ would be the diffusion operator. To leading semiclassical
order the correlation function then assumes the form $R_2(\omega) =\frac{1}{2}
\left(\frac{\Delta}{\pi}\right)^2
\sum_{x_+,x_-}\Pi(x,x',\omega)\Pi(x_-,x_+,\omega)$ of a one-loop diagram (inset  left) involving the temporal Fourier transforms of the mode propagators. 

The solution of the propagator master equations crucially depends on the symmetries
and conservation laws of the underlying scattering processes. For example, the
averaging over a single particle random potential effectively restores translational
invariance meaning that the difference in momenta, $q$, between the participating
states is conserved (inset top). The conserved momenta play a role of effective
quantum numbers of the mode propagators, and at the same time are Fourier conjugate
to the coordinate difference $x-x'$ in $\Pi(x,x',t)$. Indeed, the diffusion operator
$ {\cal D} \partial_x^2$ is diagonal in a momentum representation and the frequency
representation of the mode equation has the solution
$\Pi(q,\omega)=-(i\omega- {\cal D} q^2)^{-1}$. With this result, the spectral two-point
function assumes the role of a sum over relaxation modes~\cite{Altshuler:1986},
\begin{align}   
    \label{eq:R2DiffusivePert}
    R_2(\omega)= \frac{1}{2} \left(\frac{\Delta}{\pi}\right)^2\textrm{Re}\sum_q \frac{1}{(i\omega- {\cal D} q^2)^2}. 
\end{align}
For frequencies larger than the Thouless energy, $\omega>E_C\equiv
t_{\textrm{erg}}^{-1}$, this expression is dominated by modes of non-vanishing
momentum, $q$. This is the non-ergodic regime affected by the diffusive kinematics of
the modes, their dimensionality-dependent density of states, etc. For smaller
frequencies, $\omega<E_C$, the sum is dominated by the momentum \emph{zero mode},
$q=0$ (unless, the sum over modes yields an UV divergent result, which in the
diffusive context would happen for $d\ge 4$. We will return to the discussion of this
situation below.) The zero mode contribution to the correlation function
$R_2(\omega)=-\frac{1}{2}\left( \frac{\Delta}{\pi \omega} \right)^2$ is fully
universal in that it depends only on the dimensionless ratio of energy difference and
single particle level spacing. For frequencies larger than the single particle level
spacing, $\Delta\ll \omega\ll E_C$, this expression agrees with the RMT result (we
assume absence of time reversal here, such that the relevant ensemble is the Gaussian
Unitary Ensemble, GUE)
\begin{align*}
    R_{2,\textrm{RMT}}(\omega)=\Delta\, \delta(\omega) -\left( \frac{\sin(\pi
\omega/\Delta)}{\pi\omega/\Delta} \right)^2\stackrel{\omega\gg \Delta}\simeq -
\frac{1}{2}\left( \frac{ \Delta}{\pi\omega} \right)^2.
\end{align*}
 We finally note that
the IR divergence in the  zero mode contribution $\sim \omega^{-2}$ at small values
$\omega<\Delta$ is cut by the emergence of `nonlinearities' in the theory. The
semiclassical precursor of these processes are higher order loop diagrams, as
indicated in the bottom inset, right. However, the non-perturbative nature of the
non-linearity (i.e. the impossibility to capture them by diagrammatic resummation) is
indicated by the non-analyticity of the RMT $\sin$-function in $1/\omega$, i.e. in
the propagator amplitude of the semiclassical zero-mode. If one is ambitious to
describe spectral statistics for all frequency values the semiclassical formulation
needs to be integrated into a field theoretical framework. In this way one finds that~\cite{Andreev:1995} 
Eq.~\eqref{eq:R2DiffusivePert} generalizes to 
\begin{align}   
    \label{eq:R2DiffusivePert}
    R_2(\omega)=R_{2,\textrm{RMT}}(\omega)+\frac{1}{2} \left(\frac{\Delta}{\pi}\right)^2\textrm{Re}\sum_{q\not=0} \frac{1}{(i\omega- {\cal D} q^2)^2}.
\end{align}
In this expression, the perturbatively singular contribution of the ergodic mode $\sim \omega^{-2}$ is regularized and absorbed in the RMT-contribution $R_{2,\textrm{RMT}}$.

\begin{figure}[t]
\centering{\includegraphics[width=6.0cm]{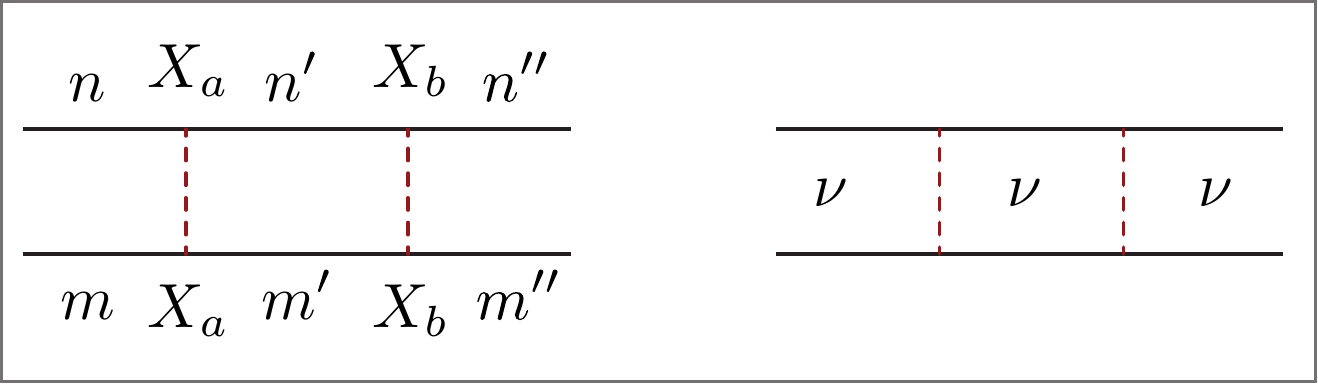}}
\caption{Left: building blocks of the Majorana relaxation modes. Scattering now is off the operators $X_a\equiv \chi_{i_1}\chi_{i_2}\chi_{i_3}\chi_{i4}$ entering the scattering Hamiltonian. Modes in Fock space are  deconfined in that the many body states $|n\rangle$ and $|m\rangle$ correlated by scattering can be very different $|n-m|=\mathcal{O}(N)$. (Here, $|n-m|$ is the Hamming distance between $n$ and $m$, i.e. the number of binary symbols in $n$ that need to be switched to get to $m$.)  Right: the conserved quantum numbers of the process are the labels, $\nu$, of the basis states of the Majorana Clifford algebra, as discussed in the text.}
\label{fig:MajoranaSpectralStatistics} 
\end{figure}

We now discuss how the general concepts introduced above carry over to the case of
the two-body Majorana scattering operator Eq.~\eqref{eq:Hamiltonian}. First note that the Hamiltonian
conserves fermion parity and commutes with the parity operator $
P\equiv\prod_{i=1}^{N/2-1}(i \chi_{2i-1}\chi_{2i})$, where we consider even values of
$N$ for definiteness. For definiteness, we will focus on systems of state number
$N=2,6,10,14,\dots$ with $N \,\textrm{mod}\, 8=2$ or $6$, which fall into the unitary
symmetry class.\footnote{The Hamiltonian Eq.~\eqref{eq:Hamiltonian} is time reversal
and particle hole symmetric under an implementation of these symmetries discussed in
detail in Ref.~\cite{You:2016}. For $N \,\textrm{mod}\,8=2,6$ the parity operator $P$ effectively
anti-commutes with the relevant anti-unitary symmetry, which means that time reversal
and particle-hole symmetry no longer are effective symmetries within the irreducible
sectors of definite parity; the Hamiltonian $H$  acts as a random hermitean but
otherwise symmetry-less operator.} The Hamiltonian acts within Hilbert space sectors of definite parity, and we will consider the $D\equiv 2^{N/2-1}$ dimensional Hilbert space $V$ of even occupation number throughout. This space is generated by the action of an even number of Majorana operators on the vacuum.  

In this setting, the role of the position states, $|y\rangle$, is taken by states
$|n\rangle\equiv|n_0,\dots,n_{N/2}\rangle$ where $n_i=0,1$ is the occupation number
of the fermion $c_i\equiv \frac{1}{2}(\gamma_{2i-1}+i\gamma_{2i})$ and $|n|\equiv \sum_i n_i$ is even. Throughout, we
will label products of Majorana operators $X_\mu\equiv
\chi_{\mu_1}\chi_{\mu_2}\dots \chi_{\mu_l}$, $l$ even, acting in $V$ by the container symbol  $\mu=(\mu_1,\dots,\mu_l)$, where
$\mu_1<\mu_2<\dots<\mu_l$ defines an ordered string of numbers $1\le \mu_i\le N$. For convenience, we add to this set the unit operator $X_0\equiv \Bbb{I}$.
Note that the operators $X_\mu$ commute or anti-commute amongst themselves, $X_\mu X_\nu=s(\mu,\nu)X_\nu X_\mu$, where the sign factor 
$s(\mu,\nu)=\pm 1$ will play a very important role throughout.
 Finally, we reserve the symbol
$X_a$, $a=(a_1,a_2,a_3,a_4)$  for the $n\equiv \left( {N\atop 4} \right)$ operators of norm $l=4$ featuring in the interaction Hamiltonian.

In this language, the propagators $\langle n_+|(\epsilon^\pm -\hat
H)^{-1}|n_+\rangle$ can be considered as sums over closed loop scattering
paths during which states $n$ scatter to states $n'$ via matrix elements
$\left\langle n'|H |n\right\rangle=\sum_a J_a \left\langle n'|X_a |n\right\rangle$ (Fig.~\ref{fig:MajoranaSpectralStatistics},
left.) A first essential difference to the previously discussed case is that the two
amplitudes correlated by a scattering sequence in Fock space can be far apart. The reason is that even for very different states $|n\rangle,|m\rangle$, the product of matrix elements $\langle\langle n'|J_a X_a|n\rangle \langle m|J_a
X_a|m'\rangle\rangle\sim \langle n'|X_a|n\rangle \langle m|X_a|m'\rangle $ may be
non-vanishing. This is to be compared to the case of single particle scattering where the amplitudes forming the particle-hole scattering channel were close to each other on scales $\sim \xi$. The `deconfined' nature of the scattering channel can be seen as a consequence of the sparsity of the random matrix $H$. 

We  now consider the mode evolution under this type of scattering dynamics. At any
instance of time, the state of the composite mode is encoded in the amplitudes
$\Pi(n,m|n_0,m_0,t)\equiv \Pi(n,m)$, where the initial configuration, $(n_0,m_0)$, and the time argument, $t$, are  suppressed in the second representation for better
readability. The state of the mode after a scattering event off the operators $X_a$
is given by $\Pi(n',m')=\sum_{n,m} \langle n'|X_a|n\rangle \langle m|X_a |m'\rangle
\Pi(n,m)$. One may write this in an index-free notation as $\Pi\to \Pi'\equiv X_a \Pi
X_a$, where $\Pi\equiv  \sum_{n,m}   \Pi(n,m) \,|n\rangle\langle m| $ is considered as
a matrix in $V$ or, equivalently, as an element of the tensor product of the Hilbert
space and its dual $V\otimes V^\ast$ . In view of this Fock space non-locality of
these objects it is all the more important to identify `mode quantum numbers'
conserved by the scattering channels. 

As in the single particle problem, progress is made by representing the particle-hole
dyads $ |n\rangle\langle m| \in V\otimes V^\ast$ in a basis different from the
original one, i.e. by identifying the SYK-analog of a `momentum representation'. To
this end let us consider $V\otimes V^\ast$ as the  $D^2$ dimensional space of Hilbert space matrices. Above, we
have introduced a specific set of elements of this space, namely the operators $X_\mu$. The action of all possible products of Majorana operators on the vacuum generates all of $V$, which is to say that the $X_\mu$ form a complete set  and that any other matrix $A\in V\otimes
V^\ast$ can be expanded as
\begin{align}
    \label{eq:OperatorExpansion}
    A=\frac{1}{D^{1/2}}\sum_\mu a_\mu X_\mu,\qquad a_\mu=\frac{1}{D^{1/2}}\textrm{tr}(A X^\dagger).
\end{align}
Here, the identification of the expansion coefficients follows from $\textrm{tr}(X_\mu
X_{\nu}^\dagger)=D\delta_{\mu,\nu}$, i.e. the fact that the trace over
non-vanishing monomials of Majorana operators vanishes, and only $X_\mu
X^\dagger_\mu=1$ has a finite trace $D$. The sum extends over one half of the 
$2^{N-1}$  even parity operators $X_\mu$: operators $X_\mu $ and $X_\nu\equiv X_\mu P$ are identified because $P$ acts as the identity
operator in $V$. For example, for $N=4$, the even operators $\chi_1\chi_2$ and $
\chi_1\chi_2 (\chi_1\chi_2\chi_3\chi_4)\propto \chi_3\chi_4$ are identical in the even subspace. This leaves  $2^{N-2}=D^2$ independent operators which form a \emph{basis} of $V\otimes V^\ast$.

Notice the similarity of the expansion Eq.~\eqref{eq:OperatorExpansion}
expansion with a Fourier transform. Indeed, we will observe that the states $\mu$
assume a role very similar to the momentum states of the single particle theory. 
Specifically, the transformation to $\mu$-states may be applied to represent the modes as
$\Pi=\frac{1}{D^{1/2}}\sum_\mu \pi_\mu(t) X_\mu$. The key observation now is that the
scattered mode, $\Pi'$, has the expansion coefficients, $\pi'_\nu=\frac{1}{D^{1/2}}
\textrm{tr}(\Pi'X^\dagger_\nu)=\frac{1}{D^{1/2}} \textrm{tr}(\Pi X_a X^\dagger_\nu
X_a)=\frac{1}{D}\sum_\mu \pi_\mu \textrm{tr}(X_\mu X_a X^\dagger_\nu
X_a)=\frac{1}{D}\sum_\mu \pi_\mu s(a,\mu) \textrm{tr}(X_\mu
X^\dagger_\nu)=s(a,\nu)\pi_\nu$. This construction shows that the scattering
sequence \emph{conserves} the $\nu$--state. Individual scattering events merely generate a sign factor
$s(a,\nu)$. We conclude
that the labels $\mu$ play a role analogous to the conserved momenta, $q$, of the single
particle problem.

In the next section we will show that the dispersion of these modes is determined by the relation $\Pi(\nu,\omega)=(i\omega- \epsilon(|\nu|))^{-1}$, where the $\epsilon(k)$ depends only on the state norm $k\equiv |\nu|$. For small values $k\ll N$ the dispersion is approximately linear $\epsilon(k)=\pi^{-1}\Delta D\left(8 k/N +\mathcal{O}(k/N)^2\right)$, and it approaches values of $\Delta D\times \mathcal{O}(N^2)$ for modes with generic value  $k\simeq N$  (see Eq.~\eqref{eq:Dispersion} for the exact expression.) Here, 
\begin{align}
    \Delta \equiv \frac{\pi J N^{1/2}}{2 D},
\end{align}
is the mean level spacing at the band center. Since $\Delta D$ is of the order of the
many body band width, generic modes are  heavy and essentially non-dispersive.
However, the gap of the lightest massive mode
\begin{align}
     \gamma_* \equiv \epsilon(2)= 16 \Delta D/(\pi N),
 \end{align} 
is much smaller and it sets the inverse of the time scale at which the longest lived
 structured modes, $k=2$, have relaxed. In this regard it plays a role analogous to
 the Thouless energy of a disordered single particle system. However, in view of the
 exponentially large density of states $\left( {N\atop k} \right)$ of modes with gap
 $\epsilon(k)$, care must be exercised in transferring results from single particle
 spectral statistics to the present context. Specifically, we observe that, unlike
 with the single particle system, corrections to RMT spectral statistics are
 observable at frequencies much lower than $\gamma$.

\begin{figure}[t]
\centering{\includegraphics[width=8.0cm]{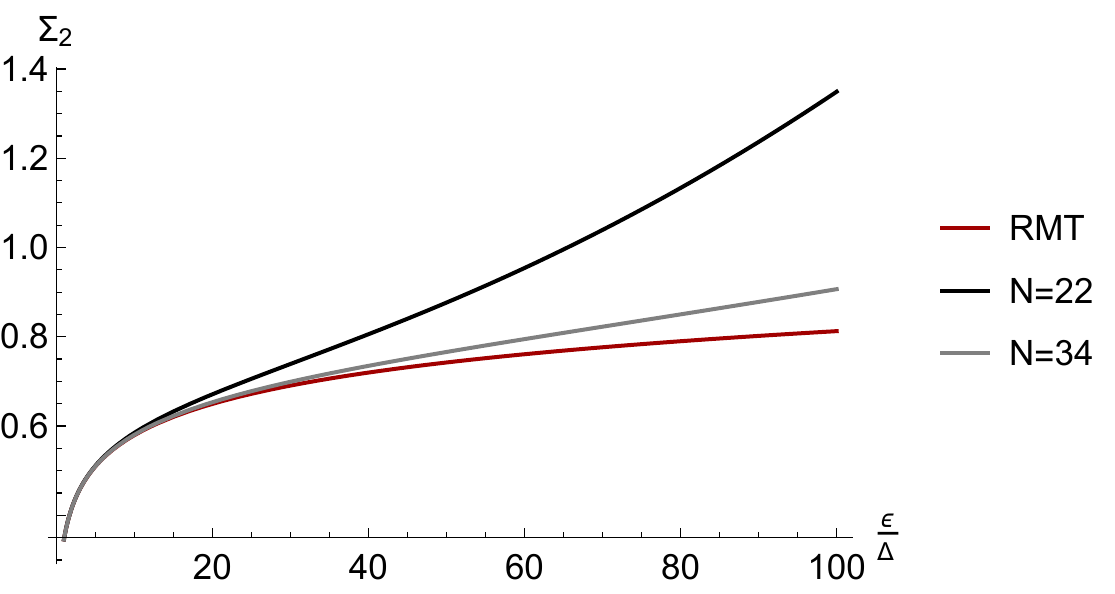}}
\caption{Number variance of systems with $N=22$ and $N=34$, resp., compared to that of the GUE. Discussion, see text.}
\label{fig:Sigma2} 
\end{figure}

To better understand these structures, we consider the effect of non-ergodic modes on various spectral correlation functions. According to the general principle discussed above, we expect the two-point correlation function to assume the form
\begin{align}
    \label{eq:R2Modes}
      R_2(\omega)=R_{2,\textrm{RMT}}(\omega)+\frac{1}{2} \left(\frac{\Delta}{\pi}\right)^2 \textrm{Re}\sum_{k\not=0,\textrm{even}} \left( {N\atop k} \right)  \frac{1}{(i\omega- \epsilon(k))^2},
  \end{align}  
which in the present context is a sum over an exponentially large number of short
lived modes. (In the main part of the paper, this expression will be derived from a replica field integral formalism.) To obtain a crude frequency-dependent criterion for the influence of the massive modes, we ask when their total contribution becomes comparable to that of
the universal zero mode. This leads to the estimate $\omega^{-2}=\sum_{k\not=0}
\left( {N\atop k}
\right)(i\omega+ \epsilon(k))^{-2}\sim D^{2}\times (D \Delta N^2)^{-2}\sim
\Delta^{-2} N^{-4}$, where we observe that the sum over non-vanishing modes is
`UV-dominated' by the exponentially large number of generic modes with their
structureless dispersion. According to this estimate, the 
contribution of non-universal modes  masks the universal contribution 
for frequency values exceeding
\begin{align}
    \omega\sim \omega_{\textrm{erg}}\equiv \Delta N^2\ll \gamma_*.
\end{align}
Only for frequencies $\omega\lesssim
 \omega_{\textrm{erg}}$ will universal spectral statistics be observed. Note that $\omega_\mathrm{erg}$ is much smaller than the lowest relaxation gap.  As an aside,
 we mention that the spectral two-point function must satisfy the sum rule $\int
 \omega R_2(\omega)=0$ (the integral over fluctuations in the density of states
 around its mean equals zero.) While this rule appears to be violated by the near
 frequency-independent background of the generic modes, we note that
 Eq.~\eqref{eq:R2Modes} holds only for frequencies $\omega\ll \Delta D$ much smaller
 than the band width. As shown later, modifications effectively restoring the sum rule take place at larger frequencies.

In order to compare the above prediction with earlier numerical work, we consider
the number variance around some value $E$ in the bulk of the spectrum,
\begin{equation}
\Sigma_2(\epsilon)=\Delta^{-2} \!\!\!\int\limits_{E-\epsilon/2}^{E+\epsilon/2} d\epsilon_1
d\epsilon_2 R_2(\epsilon_1-\epsilon_2)=
2 \Delta^{-2}\!\int\limits_{0}^\epsilon d\omega
(\epsilon-\omega) R_2(\omega)\simeq \frac{1}{\pi^2}\ln \left(
\frac{2 \pi\epsilon}{\Delta} \right)+\frac{4 \pi \epsilon^2}{\Delta^2 N^4}.
\end{equation}
Here, the first term is derived by integration of the RMT-contribution to the spectral two point function over energy. The second term is obtained from the massive mode contribution, noting that for the exponential majority of them the dispersion ($\omega$ compared to $\epsilon(k)$) is negligibly weak.
The result is in semi-quantitative agreement with the numerical data shown in Ref.~\cite{Garcia:2016}:
The deviations from the RMT limit show a convex upturn as a function of energy, and they are stronger for smaller
$N$. An eyesight inspection of the data suggests that the deviations of the $N=22$
fluctuations exceed those of $N=34$ by a factor of about $7$. This is not far off
the above result which would predict a ratio $34^4/22^4\simeq 5.7$.

\begin{figure}
\centering{\includegraphics[width=14.0cm]{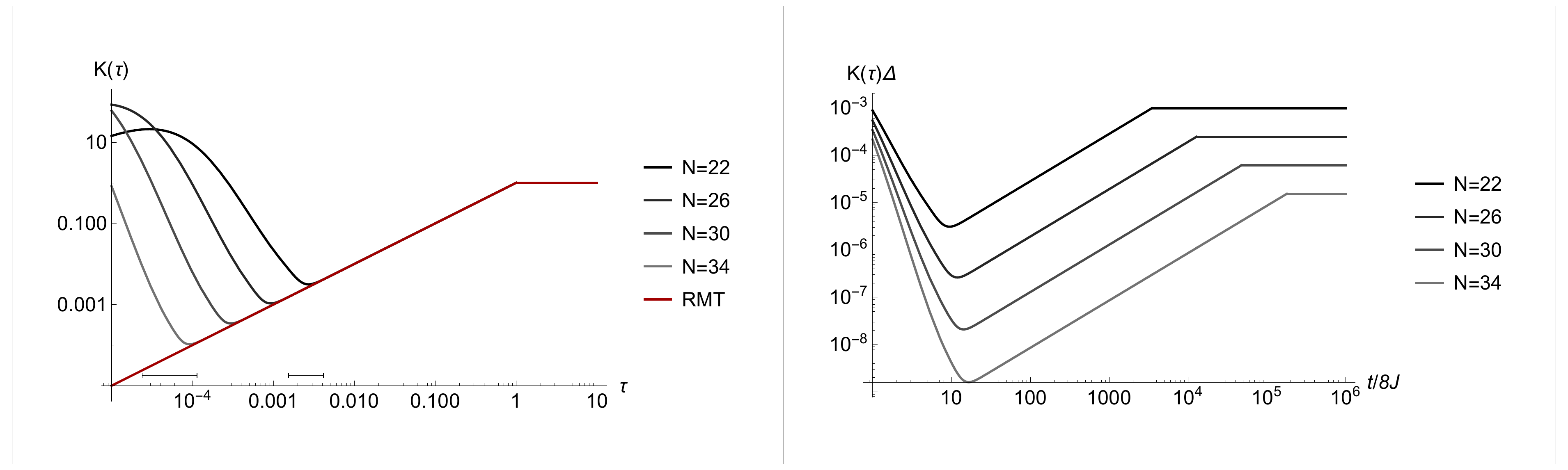}}
\caption{Spectral form factor as a function of scaled time $\tau$ (left) and of physical time $t$ (right). Discussion, see text.}
\label{fig:FormFactor} 
\end{figure}

For a  more structured comparison we now turn to the discussion of the spectral form factor. The straightforward Fourier transform of the two-point function leads to 
\begin{align}
    \label{eq:SpectralFormFactor}
      K(\tau)&=K_{\mathrm{RMT}}(\tau)+\tau\sum_{k\not=0, \mathrm{even}}\left( {N\atop k} \right) 
e^{-\tau \frac{2\pi\epsilon(k)}{\Delta}}, \quad \tau>0.
\end{align}   
Here, $K_{\mathrm{RMT}}(\tau)=\tau \Theta(1-\tau)+\Theta(\tau-1)$ is the RMT form
factor obtained by Fourier transformation of the non-perturbative zero mode
contribution $R_{2,\mathrm{RMT}}$. The sum represents the contribution of non-ergodic
modes. It is multiplied by a factor $\tau$ safeguarding the limit $K(\tau\to 0)\to 0$
required by unitarity (i.e. by the sum rule $\int d\omega R_2(\omega)=0$ describing
the constancy of the total number of levels.) However, we emphasize that the
result~\eqref{eq:SpectralFormFactor} is based on an effective low energy theory which
looses validity for energies $\omega\sim D \Delta$ of the order of the band width,
corresponding to dimensionless times $\tau\lesssim \tau_\mathrm{UV}\sim D^{-1}$. This means that the quantitative form by which
$K(\tau)$ of Eq.~\eqref{eq:SpectralFormFactor} approaches zero for times shorter than
$\tau_\mathrm{UV}$ must not be taken seriously.

The left panel of Fig.~\ref{fig:FormFactor} shows the form factor for the four values
$N=22, 26, 30$ and $34$, respectively, compared to that of an RMT ensemble. At an
$N$-dependent time, $\tau_\mathrm{erg}$, the form factor exhibits a pronounced
minimum and for larger times approaches the RMT result. A straightforward variational
computation shows that the minimum is located at $\tau_{\mathrm{erg}}\sim \ln(N) N
D^{-1}$. This is only by a factor of $\sim N \ln(N)$ larger than short UV time
cutoff, $ \tau_{\mathrm{UV}}$ (for $N=22$ and $34$ the time span between the two scales is indicated by
the horizontal bars in the left panel of Fig.~\ref{fig:FormFactor}).  At the same time, $\tau_{\mathrm{erg}}$ is the
time below which deviations off RMT behavior become visible \emph{in the form
factor}. That this time is not in straightforward inverse relation to the energy
$\omega_{\mathrm{erg}}$ above which deviations off RMT behavior become strong in the
spectral two-point function has to do with the fact that in either case the
deviations are caused by a very large number of very short lived modes. 
Where these modes
give a largely structure-less contribution to $R_2(\omega)$, their
fast relaxation in time means that they   are not  felt at
times larger than $\tau_\mathrm{erg}$ in the form factor $K(\tau)$; 
RMT correlations are better visible in $K(\tau)$ than in $R_2(\omega)$.

The profiles shown in Fig.~\ref{fig:FormFactor}
superficially resemble the `dip-ramp-structures' discussed in Ref.~\cite{Cotler:2017}. That reference considered correlations in a quantum partition sum $Z(z)\equiv
\mathrm{tr}(\exp(-z H))$  generalized to complex
`temperatures', $z$. Specifically, it considered the time-dependent correlation function $g_c(t,\beta)\equiv \langle
Z(\beta+it)Z(\beta-it)\rangle_c /\langle Z(\beta)Z(\beta)\rangle$, where $\beta$ is physical temperature. For finite $\beta$, this function, likewise  termed `form factor'
in Ref.~\cite{Cotler:2017}, is different from the scaled spectral form factor
$K(t)\equiv K(\tau)|_{\tau=2\pi t \Delta}$.  In particular,  $K(t)$ has the
limiting behavior $K(\tau\to 0)\to 0$,  whereas $g_c(t)$ asymptotes to a finite value.   
For finite $\beta$, the deviations between $K(t)$ and the thermal correlation function
$g_(t,\beta)$ essentially originate in their different short time asymptotics.
However, for the case $\beta=0$ the functions coincide, and a direct comparison to
the numerical data shown in Fig. 12 of Ref.~\cite{Cotler:2017} is possible. To ease this comparison,
the left panel of Fig.~\ref{fig:FormFactor} shows the form factor $K(\tau) \Delta $ as a
function of dimensionless time $t8/J$ for  four system sizes, $N=22, 26, 30, 34$, of unitary
symmetry included in the numerical analysis. A  comparison of the curves indicates
that the analytical calculation reproduces the essential features of the $N$-dependent dip-ramp profile seen in the numerical data. There are quantitative deviations by numerical factors of $\mathcal{O}(10\%)$ of the absolute values of minima and ramp positions. However, by and large the comparison looks favorable and we conjecture that the dip-profile is caused by the non-ergodic  modes discussed above.

In the next section we will discuss how the modes generating the spectral correlations of the system emerge as effective low energy degrees of freedom of a replica field theory.

\section{Replica field theory} % (fold)
 \label{sec:replica_field_theory}
 
 The functions $R_2(\omega), \Sigma(E)$ and $K(\tau)$ discussed above are all
 obtained from the correlation function, $C(\omega)$, Eq.\eqref{eq:CDef}. In this
 section we derive a replica generating functional, $Z(h)$, from which $C$ is
 obtained by differentiation. Compared to other approaches, the principal difference
 is that we view the problem from a \emph{first} rather then second quantized perspective. In this way of thinking, the Hamiltonian $H$ is considered as a
 large sparse random matrix acting in a $D$-dimensional
 Hilbert space and its resolvents $G^\pm (E)=(E-H)^{-1}$ are obtained from a Gaussian integral (rather than a field integral)
 \begin{align*}
     \mathrm{tr}(G^+(E))= \partial_{h^+}\lim_{R\to 0}\frac{1}{R}\int D(\bar \psi,\psi) \, \exp(-\bar \psi (E+h^+  +i\delta-H)\psi).
 \end{align*}
  Here,
 $\psi=\{\psi_n^r\}$ is a $R\cdot D$-component vector of Grassmann variables carrying
 indices $n$ in Hilbert space and $r\in \{1,\dots,R\}$ in replica space and we use the shorthand $\partial_{h^+}\equiv \partial_{h^+}|_{h^+=0}$. We will
 suppress these indices when possible, e.g., $\bar \psi (E+h^+
 +i\delta-H)\psi=\sum_r \psi^r(E+h^+  +i\delta-H)\psi^r$. The above relation
 follows from the fact that the Gaussian integration over a Grassmann field yields
 the determinant of the corresponding matrix kernel, i.e. $G^+(E)=
 \partial_{h^+}\lim_{R\to 0}\frac{1}{R}\det((E+h^+  +i\delta)-H)^R=
 \partial_{h^+}\mathrm{tr}\ln((G^+(E))^{-1}+h^+)$.

 Multiplying this with the analogous relation for $G^-$, we obtain
 \begin{align}
    \label{eq:CStartingRepresentation}
     C(\omega)&=\partial^2_{h^+ h^-}\lim_{R\to 0}\frac{1}{R^2}Z(h),\qquad\cr
     Z(h)&=\int D(\bar \psi,\psi)\left\langle\exp(- \bar \psi (\hat z-H)\psi)\right\rangle,
 \end{align}
where $\psi=\{\psi^{r,s}_n\}$ is now a $2\cdot R\cdot D$ dimensional field comprising
an index $s=\pm$ distinguishing between advanced and retarded Green functions, $\hat
z = \epsilon +(\frac{\omega}{2}+i \delta)\tau_3 +\hat h$ is a $2\times 2$ matrix in
advanced/retarded space comprising energy arguments and sources $\hat h\equiv
\mathrm{diag}(h^+,h^-)$, and $\tau_i$ are Pauli matrices acting in the same space.

We may now perform the Gaussian average over randomness to obtain
\begin{align}
     Z(h)&=\int D(\bar\psi,\psi)\, e^{-\bar \psi\hat z\psi+
\frac{3 J^2}{N^3}\sum_a (\bar \psi  X_a\psi) (\bar \psi X_a\psi)}=\cr
     &=\int D(\bar\psi,\psi)\, e^{-\bar \psi\hat z\psi-
 \frac{3 J^2}{N^3} \sum_a \mathrm{Tr}((\psi \, \bar \psi\,  X_a)(\psi \, \bar \psi\, X_a))},
 \end{align} 
where in the second step we have rearranged the quartic term from a scalar product in
$\psi$ to a dyadic product, and $\mathrm{Tr}\equiv \mathrm{tr}_V \mathrm{tr}$ is a
trace over both, Hilbert space, $V$, and the $2R$-dimensional internal space of the
$\psi$-state. This way of rewriting the nonlinearity is advantageous because the
dyads $\psi_n^+ \bar \psi_{n'}^- X_a$ are the precursor building blocks of the
two-particle modes shown in Fig.~\ref{fig:MajoranaSpectralStatistics}. Following
standard procedures, we decouple the quartic term with a Hubbard-Stratonovich transformation and integrate over Grassmann fields to obtain
\begin{align}
    Z(h)=\int DA \, e^{\frac{1}{2n}\sum_a\mathrm{Tr}(X_aA_a)^2+\mathrm{Tr}\ln\left(\hat z+
\frac{\gamma}{n}\sum_a  A_a\right)}.
\end{align}
Here, $A_a=\{A^{rr',ss'}_{a,nn'}\}$ are $n=\binom{N}{4}$ matrices of dimension $2DR$ containing complex commuting variables
and the energy scale $\gamma$ is defined as
\begin{equation}
\gamma =\frac 12 J N^{1/2}.
\end{equation}
Implicit to the definition of the transformed $Z(h)$ is a constraint on the integration variables that guarantees the  existence of the integral. 

At this point it becomes advantageous to switch to the $\mu$-representation of
operators: we define $A_a=\frac{1}{D^{1/2}}\sum_\mu a_{a,\mu} X_\mu$ where the
coefficients are $2R$-dimensional matrices $a_\mu=\{a^{rr',ss'}_\mu\}$ in the
internal indices. Using the relations discussed in the previous section we then
obtain
\begin{align}
    \mathrm{Tr}(A_a X_a A_a X_a)=\frac{1}{D}\sum_{\mu,\nu}\mathrm{tr}(a_{a,\mu}a_{a,\nu})\mathrm{tr}_V(X_\mu X_a X_\nu X_a)=\sum_\mu \mathrm{tr}(a_{a,\mu}a_{a,\mu})s(\mu) s(a,\mu).
\end{align}
We here defined
\begin{align}
    X_\mu \equiv X^\dagger_\mu s(\mu),
\end{align}
where $s(\mu)=s(|\mu|)$ is a sign factor depending only on the number of Majoranas contained in $\mu$. (Straightforward combinatorics shows that $(s(0),s(1),s(2),s(3),s(4),\dots)=(+,+,-,-,+,\dots)$, or $s(n)=(-)^{\lfloor (n/2)\rfloor}$, where $\lfloor x\rfloor$ is the floor function, i.e. $\lfloor 1\rfloor=\lfloor 3/2\rfloor=1$, etc. However, we will not need this explicit definition in the following.) The combination of sign factors appearing in the sum above follows from
\begin{align}
    \mathrm{tr}_V(X_\mu X_a X_\nu X_a)=\mathrm{tr}_V(X_\mu X^2_a X_\nu)s(a,\mu)=\mathrm{tr}_V(X_\mu  X_\nu)s(a,\mu)=\mathrm{tr}_V(X_\mu^\dagger  X_\nu )s(a,\mu)s(\mu)=\delta_{\mu,\nu}s(a,\mu)s(\mu),
\end{align}
where we used that quartic products of Majoranas square to unity, $X_a^2=1$. The advantage of the new representation is that the trace of the Gaussian weight has collapsed to one over the internal indices. Notice the diagonality of the weight in the Hilbert space indices, $\mu$, which is based on a construction identical to that  demonstrating the $\mu$-conservation of the two-particle scattering vertex. Indeed, the Hubbard-Stratonovich matrices $A_{a,nn'}\sim \psi_n \bar \psi_{n'}X_a$ represent bilinears of particle amplitudes decorated by scattering vertices, and $a_{a,\mu}$ are these bilinears in the $\mu$-representation.  

We now observe that the non-linearity tr ln of the integral couples only to the homogeneous configuration $\bar A\equiv \frac{1}{n}\sum_a A_a$. This suggests a shift, $A_a\to \bar A + A_a$, where a constraint $\sum_a A_a=0$ for the shifted variables is understood. The same change of variables is applied to the $\mu$-variables,  $a_{a,\mu}\to\bar a_\mu+ a_{a,\mu}$, with  $\sum_a a_{a,\mu}=0$. Using this representation, the functional integral becomes
\begin{align}
    \label{eq:ZBeforeIntInh}
    Z(h)=\int D(a,\lambda) \, e^{\frac{1}{2n}\sum_a\mathrm{tr}(\bar a_{\mu}+a_{a,\mu}))^2s(\mu) s(a,\mu)+\frac{1}{n}\sum_{a,\mu}s(\mu) \mathrm{tr}(\lambda_{\mu} a_{a,\mu})+
    \mathrm{Tr}\ln\left(\hat z+\gamma \bar A\right)},
\end{align}
where $\lambda_{\mu}$ are Lagrange multiplier matrices implementing the constraint. The integrations over $a_{a,\mu}$ are now Gaussian and can be carried out in closed form. As a result of a straightforward procedure detailed in \ref{sec:DerInhomogeneousmodesIntegrated}, we obtain the functional integral $Z(h)=\int Da\exp(-S[a])$
\begin{align}
    \label{eq:ActionInhomogeneousModesIntegrated}
    S[a]=-\frac{1}{2}\sum_\mu s(\mu) S(\mu)^{-1} \mathrm{tr}(a_\mu^2) - 
\mathrm{Tr}\ln\left(\hat z+\frac{\gamma}{D^{{1/2}}}\sum_\mu a_\mu X_\mu\right),
\end{align}
where we omitted the overbar, $\bar a\to a, \bar
A\to A$ for notational brevity and defined
\begin{align}
    S(\mu)\equiv \frac{1}{n}\sum_a s(a,\mu).
\end{align}
Notice that while the Gaussian weight $\sim n^{-1}\sim N^{-4}$ of the matrices $a_{a,\mu}$ was `light', that of the modes $a_\mu$ is heavier. For generic $\mu$, $|\mu|=\mathcal{O}(N)$, the number of operators $X_a$ commuting/anti-commuting with $X_\mu$ is roughly equal implying that $S(\mu) \sim \frac{\sqrt{n}}{n}\sim N^{-2}$ and the Gaussian weight scales as $\sim N^{2}$. 

\section{Stationary phase analysis} % (fold)
\label{sec:StationaryPhaseAnalysis}

In this section we subject the effective action to a stationary phase analysis. The legitimacy of the procedure will be checked self-consistently at a later stage. A variation of the action Eq.~\eqref{eq:ActionInhomogeneousModesIntegrated} over $a_\mu$ yields the stationary phase equation
\begin{align}
    \label{eq:MeanFieldEq}
    s(\mu) S(\mu)^{-1} \frac{a_\mu}{D^{{1/2}}} +  
\gamma\frac{1}{\hat z+ \gamma\sum_\nu X_\nu \frac{a_\nu}{D^{ {1/2} }}}X_\mu=0.
\end{align}
For energies, $E$, in the center of the band, this equation is solved by a Hilbert space homogeneous ansatz, 
$a_\mu\equiv D^{{1/2}}\hat y\delta_{\mu,1}$. For this configuration, $S(\mu)=S_1=1$, and the equation reduces to 
\begin{align*}
    \hat y+\gamma\frac{1}{\hat z+\gamma \hat y}=0.
\end{align*}
This is a quadratic equation and it is solved by
\begin{align}
    \label{eq:MeanFieldSolution}
      \hat y= -\frac{\hat z}{2\gamma } + i\tau_3 \sqrt{1-\frac{\hat z}{2\gamma}}\,\sqrt{1+\frac{\hat z}{2\gamma}},
 \end{align} 
 where we noted that the sign of the square root is determined by the imaginary part of $\mathrm{Im}(\hat z)=\delta \tau_3$.
Substitution of this solution leads to the mean field action
\begin{align*}
S[\hat y]=
- \frac{D}{2} \mathrm{tr} \left( \frac{{\hat z}^2}{2\gamma^2} - i \tau_3 \frac{\hat z}{\gamma}
  \sqrt{1-\left(\frac{\hat z}{2\gamma}\right)^2} \right)
- \mathrm{Tr}\ln \left( \frac{\hat z}{2\gamma} + i \tau_3 \sqrt{1-\left(\frac{\hat z}{2\gamma}\right)^2} \right).
\end{align*}
%where we noted that the evaluation of the Gaussian action on the mean field yields an inessential constant, vanishing in the replica limit. 
If we differentiate \emph{once} w.r.t. sources and set $\omega=0$, we obtain the mean field estimate for the average density of states,
\begin{align}
    \rho(\epsilon)=-\frac{1}{\pi}\mathrm{Im}\,\mathrm{tr}(G^+(\epsilon))=
    -\frac{1}{\pi}\mathrm{Im}\,\partial_{h^+}Z(h)= \frac{D}{\pi\gamma}\sqrt{1-\left( \frac{\epsilon}{2\gamma} \right)^2}.
\end{align}
This formula states that (i) the average level spacing in the band center is given by
\begin{align}
    \Delta = \rho^{-1}(0)=\pi \gamma D^{-1}=\frac{\pi}{2} J N^{1/2} D^{-1} ,
\end{align}
(ii) the characteristic many body band width is given by
\begin{align}
    \Gamma\equiv 2\gamma = J N^{1/2},
\end{align}
(implying that $\Delta \sim \Gamma /D$), and (iii) at the level of the above mean
field approximation, the density of states in the bulk of the band is given by a
semicircular profile. It is well known~\cite{Garcia:2017}, that the last statement is approximate. The
density of states even in the bulk of the band is better approximated by a Gaussian,
and in the tails approaches a square root dependence. The above  solution of the self
consistent Born type equation~\eqref{eq:MeanFieldEq} can be made more accurate by the
combinatorial methods of Ref.\cite{Garcia:2017}. Close to the band edges, corrections become strong
and a full solution of the problem~\cite{Bagrets:2017} leads to the many body density of states
computed by different methods in Refs.~\cite{Cotler:2017, Stanford:2017}. However, in the present context we are
primarily interested in the \emph{correlations} of the DoS at nearby energies and the
weak dependence of the average DoS on the center energy $\epsilon$ is of secondary
importance. For this purpose the semicircular estimate~\eqref{eq:MeanFieldSolution}
is good enough.

\section{Fluctuations (RMT)} % (fold)
\label{sec:FluctuationsRMT}

We now turn to the discussion of fluctuations around the mean field and their
ramification in spectral statistics. To begin with, note that in the limit $\omega\to
0$ the starting functional Eq.~\eqref{eq:CStartingRepresentation} is invariant under
transformations $\psi\to T\psi$, $\bar \psi \to \bar \psi T^{-1}$ homogeneous in
Hilbert space, $T=\{T^{rr',ss'}\}$. The action thus possesses a $G\equiv \mathrm{GL}(2R,2R)$
replica symmetry, weakly broken by $\omega$. The mean field solution
(spontaneously) breaks this symmetry down to $H\equiv \mathrm{GL}(R,R)\times
\mathrm{GL}(R,R)$, i.e. the transformations commuting with $\tau_3$. As a result of
this symmetry breaking a coset space $G/H$ of Goldstone modes appears. In the context
of single particle physics, these Goldstone mode fluctuations are the degrees of
freedom of the  nonlinear sigma model approach to disordered systems. Their
appearance is made explicit by noting that the mean field
equation~\eqref{eq:MeanFieldEq} possesses the continuous manifold of solutions
 $\gamma T \hat y T^{-1}\simeq  - \frac{\epsilon}{2}+\frac{i\Gamma}{2}Q$, where $Q=T\tau_3 T^{-1}$.
The fluctuations $T$ are soft modes of the theory and must be integrated
over. Substituting the fluctuation configurations into the action, noting the
invariance of the Gaussian action and expanding to first order in   $\omega/\Gamma
\ll1$ we obtain
\begin{align}
    \label{eq:SSigmaModel}
    S[Q]=-\mathrm{Tr}\ln \left( \hat z + \gamma T\hat yT^{-1}  \right)= -\mathrm{Tr}\ln \left( T^{-1}\hat z T +\gamma \hat y  \right)\simeq \frac{i\pi}{\Delta} \mathrm{tr}(Q\hat z),
\end{align}
where in the last step we used that $\frac{1}{\gamma \hat y}\simeq-\frac{i\pi}{D\Delta}\tau_3+\mathrm{const.}$, and the constant (vanishing in the replica limit may be ignored.
The expression on the left is the action of the zero-dimensional nonlinear $\sigma$-model of disordered systems. This theory is in quantitative correspondence with RMT. Specifically, the integration over $T$ leads to the RMT spectral correlation function
\begin{align}
    \label{eq:R2RMT}
    R_{2,\mathrm{RMT}}(\omega)=\Delta \delta(\omega)
-\left( \frac{\sin(\pi \omega/\Delta)}{\pi\omega/\Delta} \right)^2\stackrel{\omega>\Delta}\simeq-\frac{1}{2} \left( \frac{\Delta}{\pi\omega} \right)^2,
\end{align}
and the Fourier transform of this expression  yields the RMT form factor $K_{\mathrm{RMT}}(\tau)$ in Eq.~\eqref{fig:FormFactor}.
 Here, the $\delta$-function contribution describes the auto-correlations of individual levels. It contributes to sum rules based on the correlation function, $R_2$, but is otherwise inessential.
For the sake of completeness, we outline the derivation of this result in~\ref{sec:RMTSpectralCorrelation}.

\section{Fluctuations (massive)} 
\label{sec:FluctuationsMassive}

The asymptotic form of Eq.~\eqref{eq:R2RMT} equals the $k=0$ contribution to
Eq.~\eqref{eq:R2Modes}. Higher order contributions are proportional to the
propagators $(i\omega-\epsilon(k))^{-1}$ and must therefore be due to `massive'
fluctuations. In the following we compute the quadratic action of these modes, and in
this way identify the weights $\epsilon(k)$ determining their mass. To this end,  we
generalize the set of integration variables to $a_\mu = T(D^{1/2} \hat
y+y_\mu)T^{-1}$, where $T\propto 1_V$ are the Hilbert space singlet Goldstone modes,
and $y_\mu$ are fluctuation matrices. For  $\mu\not=0$ these modes carry structure in
Hilbert space (for completeness we note that the singlet mode $y_{0}$ is diagonal in
advanced/retarded, because off-diagonal fluctuations are already accounted for by
$T$.)  The substitution of this
ansatz into the action~\eqref{eq:ActionInhomogeneousModesIntegrated} leads to
\begin{align}
    \label{eq:GaussianActionMassive}
    S[Q,y]&=-\frac{1}{2}\sum_\mu s(\mu) S(\mu)^{-1}\,\textrm{tr}(D^{1/2}\hat y + y_\mu)^2-\mathrm{tr}\ln \left( T^{-1}\hat z T + \gamma\hat y +\frac{\gamma}{D^{1/2}}\sum_\mu X_\mu y_\mu \right)\simeq \\
    &\simeq S[Q]-\frac{1}{2}\sum_\mu s(\mu) \left\{ S(\mu)^{-1}\, \textrm{tr}(y_\mu^2 )-\textrm{tr}\left( \left[ \frac{\hat z}{2\gamma} - i \tau_3 \sqrt{1- \left( \frac{\hat z}{2\gamma} \right)^2  } \right) y_\mu  \right]^2  \right\}\equiv S[Q]+S_{\textrm{m}}[y],\nonumber
\end{align}
where we have used $X_\mu^2 = s(\mu)$ and neglected the coupling between the $y_\mu$ and the $Q$-fluctuations
 (In regimes, where the cumulative contribution of the $y_\mu$ is sizeable, $\omega\gg \Delta$ fluctuations of $Q$ are small, and $Q\simeq \tau_3$ is a reasonable approximation). The structure of the action suggests a decomposition
\begin{align}
    y_\mu = w_\mu + v_\mu,
\end{align}
in contributions off-diagonal and diagonal in advanced/retarded space, respectively. The diagonal fluctuations, $v_\mu$, describe correlations between Green functions of identical causality and do not couple to spectral correlation functions. This statement extends beyond the Gaussian order considered presently (see section~\ref{sec:FluctuationsMassiveNonlinear}) and we will therefore neglect these fluctuations throughout. A quick calculation, detailed in~\ref{sec:GaussianActionMassive} leads to
\begin{align}
    \label{eq:GaussianActionMassiveEffective}
    S_{\textrm{m}}[w]\simeq -\frac{1}{2}\sum_\mu s(\mu) \,\mathrm{tr}\, 
\biggl( \left(\Pi_\mu^{-1}-i\Delta h/\gamma \right)\, w_\mu^2\biggr),
\end{align}
where $\Delta h\equiv h_+-h_-$, and we defined the `propagators' $\Pi_\mu^{-1}\equiv
 S(\mu)^{-1}-1  - i\omega/\gamma$. Since $w_0=0$, 
the sum over $\mu$ starts with configurations containing at least two Majoranas, the lowest order non-trivial even parity configuration. 

Now it is a good time to analyze the factors
$S(\mu)=\frac{1}{n}\sum_a s(a,\mu)$ giving these modes their weight. Consider a
configuration $\mu$ containing $k=2l$ operators $\chi_i$, i.e. $|\mu|=k$. There are
$\binom{N-k}{4}$ quartic configurations, $a$, which have no Majoranas in common with
$\mu$, and hence trivially commute. $k\binom{N-k}{3}$ operators have one Majorana
with $\mu$ in common and therefore anti-commute, etc. Summing over all five variants,
we obtain
\begin{align}
    \label{eq:Dispersion}
    S(\mu) \equiv S_{|\mu|}=\frac{1}{\binom{N}{4}}\sum_{j=0}^4 \binom{k}{j}\binom{N-k}{4-j}(-)^j=1-\frac{8k}{N}+\mathcal{O}(k/N)^2,
\end{align}
where the last approximation is valid for small $k$. For configurations with generic $k=\mathcal{O}(N/4)$, $|S_{|\mu|}|\sim N^{-2}$, which follows from the fact that in this case, $S(\mu)$ sums over $\sim N^4$ quasi-random sign factors. Defining 
\begin{align}
    \epsilon(k)=\gamma\left( S_{k}^{-1}-1 \right) =\frac{8 \gamma k}{N}+\mathcal{O}(k/N)^2,
\end{align}
the propagator assumes the form $\Pi_\mu = \gamma (\epsilon(|\mu|)-i\omega)^{-1}$. In view of the positivity of this expression, the integration over the off-diagonal matrices $w_\mu$ can be made convergent if we define
\begin{align}
    w_\mu = \left( \begin{matrix}
        & b_\mu\cr  - \tilde{b}_\mu
    \end{matrix} \right),\qquad \tilde{b}_\mu \equiv  s(\mu) b_\mu^\dagger. 
\end{align}
With this parameterization, the action assumes the form
\begin{align}
    S_{\textrm{m}}[b,b^\dagger]=\sum_\mu  \,\mathrm{tr}\left(\left(\Pi_\mu^{-1}-i\Delta h/\gamma\right)b_\mu b_\mu^\dagger\right).
\end{align}
We may now perform the Gaussian integration over the $R^2$ independent complex variables parameterizing each $b_\mu$ to obtain the correlation function as
\begin{align}
    Z(h)=Z_\mathrm{RMT}(h)\prod_\mu \left(\frac{1}{\Pi_\mu^{-1}-i\Delta h/\gamma}\right)^{R^2}.
\end{align}
The substitution of this expression into Eq.~\eqref{eq:CStartingRepresentation} leads to
\begin{align}
    \label{eq:CModes}
    C(\omega)=C_{\mathrm{RMT}}(\omega)+\gamma^{-2}\!\!\!\!\sum_{\mu\not=0, \,{\rm even}}\Pi^2_\mu(\omega),
\end{align}
and from this result we obtain the spectral correlation function~\eqref{eq:R2Modes}. Eq.~\eqref{eq:CModes} is the main technical result of this paper.

\section{Fluctuations (massive, nonlinear)} % (fold)
\label{sec:FluctuationsMassiveNonlinear}
 
Eq.~\eqref{eq:CModes} was obtained by quadratic expansion of the nonlinear tr ln in the action. However, in view of the exponentially large number of modes, one may wonder what happens if the expansion is pushed to higher orders, and nonlinear couplings between \emph{different} modes enter the play. In this section we show that, perhaps surprisingly, the cumulative effect of these couplings is weak and the result above remains unaltered. While this is an essential step of the programs it is also technically the most involved and readers willing to trust us on this point are invited to jump to the conclusions.

The real `danger' emanating from higher order expansions in $y_\mu$ is a potential
renormalization of the quadratic action via the cumulative effect of the other modes,
i.e. by terms of $\mathcal{O}(D)=\mathcal{O}(2^N)$. To begin exploring possible
scenarios, we note that the expansion of the tr ln in $Y\equiv D^{-1/2}\sum_\mu y_\mu X_\mu$ leads to
terms $\textrm{Tr}\ln(Y^n)=D^{-n/2}\sum_{\mu_1,\dots,\mu_n}\textrm{tr}(y_{\mu_1}\dots
y_{\mu_n})\,\textrm{tr}_V(X_{\mu_1}\dots X_{\mu_n})=\pm
D^{-n/2+1}\sum_{\mu_1,\dots,\mu_n} \textrm{tr}(y_{\mu_1}\dots
y_{\mu_n})\delta_{\mu_1 \mu_2\dots \mu_n,0}$. Here we introduced the product notation
$\mu_1 \mu_2\dots$  for the ordered product of Majorana operators contained in
$\mu_1,\mu_2,\dots$  \emph{modulo} sign factors. For example, with $\mu=(3, 4)$ and
$\mu'=(2, 3, 5)$, $\mu\mu'=(2,4, 5)$. 
This mimics the structure of the product
$X_\mu X_{\mu'}=(\chi_3 \chi_4)( \chi_2 \chi_3 \chi_5)=- \chi_2 \chi_4\chi_5=-X_{\mu
\mu'}$. The symbol $\delta_{\mu,0}$ enforces the absence of generator symbols in
$\mu$, again up to signs. For example, with $\mu=(3,4)$ we have $\mu\mu=-1$ and
$\delta_{\mu\mu,0}=1$. The meaning of the formula $\textrm{tr}_V(X_{\mu_1}\dots
X_{\mu_n})=\pm \textrm{tr}_V(X_{\mu_1 \dots \mu_n})=\pm D \delta_{\mu_1\dots
\mu_n,0}$  is that only terms in which all Majorana combine to the unit operator
survive the trace. Once more, we observe the similarity between $\mu$ and a
`momentum' label. The relation above reflects the $\mu$-conservation of the theory,
much like the trace over a single particle Hilbert space of real space sites
conserves momentum.

We study the question whether or not nonlinear terms strongly modify the theory in the band center, $\epsilon=0$, and neglecting the small energy difference, $\omega$. In this case, we may set $\hat y=i\tau_3$, which simplifies the calculation but otherwise is inessential. 
The action then assumes the form,
\begin{align*}
    S[y]&=-\frac{1}{2}\sum_\mu s(\mu) \textrm{tr}(y_\mu^2\Pi_\mu^{-1}) -\textrm{Tr}\ln(i\tau_3 +\frac{1}{D^{1/2}}\sum_\mu y_\mu X_\mu)=\cr
    &=-\frac{1}{2}\sum_\mu s(\mu) \textrm{tr}(y_\mu^2\Pi_\mu^{-1}) +\sum_{n=3}^\infty \frac{i^n}{n}\textrm{Tr}\left( \frac{1}{D^{1/2}}\sum_\mu \tau_3 y_\mu X_\mu \right)^n, 
\end{align*}
where $\Pi_\mu =  S(\mu)^{-1} - 1$ is the zero-frequency propagator. 
To further simplify our life, we assume $y_\mu=w_\mu$, i.e. we neglect the fluctuations diagonal in advanced/retarded space. (One can convince oneself, that the two types of fluctuations do not mix in the contributions of leading order in $D$ to the perturbation expansion.) The action then assumes the form
\begin{equation}
\label{eq:Action_b_expanded}
    S[b,\tilde b]=\sum_\mu s(\mu) \textrm{tr}(b_\mu \tilde b_\mu\Pi_\mu^{-1}) +\sum_{m=2}^\infty \frac{(-D)^{-m}}{m}\sum_{\mu_1,\dots,\mu_{2m}}\textrm{tr}\left(b_{\mu_1}\tilde b_{\mu_2}\dots \tilde b_{\mu_{2m}}\right) \textrm{tr}_V(X_{\mu_1}X_{\mu_2}\dots X_{\mu_{2m}}), 
\end{equation}
where we noted that only even powers of advanced/retarded off-diagonal matrices survive the trace. 

We consider the exponentiated action expanded in $y_\mu$ and perform the Gaussian integration over $m-2$ matrices $b_\mu$ in terms of total order $\mathcal{O}(b^m)$. The result can be interpreted as a contribution to a `Hartree-Fock' renormalized quadratic action, and the question is whether large renormalization contributions are generated in this way. If not, the stability of the quadratic theory has been shown, at least perturbatively.   

\begin{figure}[t]
\centering{\includegraphics[width=10.0cm]{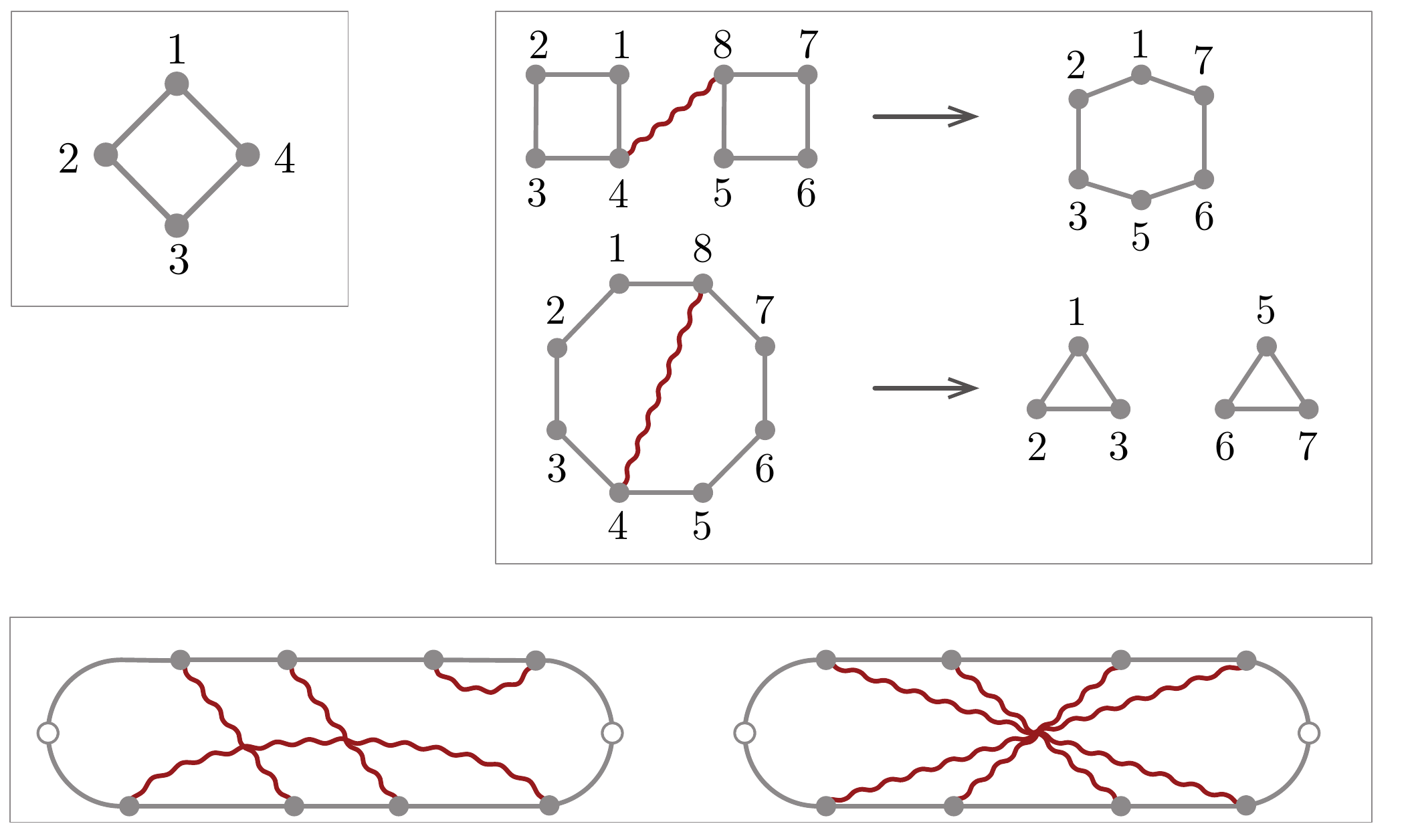}}
\caption{Contraction rules used in the perturbative computation of traces of higher order in the matrices $y_\mu$. Discussion, see text.}
\label{fig:ContractionRules} 
\end{figure}

The expansion of the action leads to expressions of the structure $\dots \textrm{tr}(b_{\mu_1}\tilde b_{\mu_2}\dots) \, \textrm{tr}(b_{\nu_1}\tilde b_{\nu_2}\dots)\dots$, i.e. products of traces of $b$-matrices. The subsequent integrals can be done by a matrix-variant of Wick's theorem. It is straightforward to verify that for fixed matrices $X,Y$ in replica space, we have the contraction rules,
\begin{align}
    \langle \textrm{tr}(Xb_\mu Y \tilde b_\nu)\rangle &=\Pi_\mu s(\mu) \delta_{\mu\nu}\textrm{tr}(X)\textrm{tr}(Y),\cr
    \langle \textrm{tr}(Xb_\mu)\textrm{tr}( Y \tilde b_\nu)\rangle&=\Pi_\mu s(\mu) \delta_{\mu\nu}\textrm{tr}(XY),
\end{align}
where the angular brackets denote the Gaussian integral over the quadratic action.
One may represent these rules graphically, as in Fig.~\ref{fig:ContractionRules},
where the $n$-corner polygons denote traces of $n$ matrices $b_{\mu_i}$ and indices
$i$ are used as an abbreviation for $\mu_i$. For example, the first upper half of the
right panel states that the contraction of $\tilde b_{\mu_4}$ and $b_{\mu_5}$ in
$\textrm{tr}(b_{\mu_1}\tilde b_{\mu_2}b_{\mu_3}\tilde b_{\mu_4})\textrm{tr}(
b_{\mu_5}\tilde b_{\mu_6} b_{\mu_7}\tilde  b_{\mu_8})$ leads to
$\textrm{tr}(b_{\mu_1}\tilde b_{\mu_2}b_{\mu_3} \tilde b_{\mu_6} b_{\mu_7} \tilde
b_{\mu_8})\Pi_{\mu_4} s(\mu_4)\delta_{\mu_4\mu_5}$, where the wavy line represents
the propagator. Note that the contraction of traces containing even powers of $y$
can generate traces of odd power, as indicated by the second panel. Finally, the
contraction of neighboring $b$'s leads to vanishing results, $\textrm{tr}(\dots b_\mu
\tilde b_{\mu}\dots)\to \textrm{tr}(\dots)\textrm{tr}(1)=\textrm{tr}(\dots)R$, which
vanishes in the replica limit. (This is the replica trick's way of eliminating vacuum
diagrams involving idle Green functions loops.)

Before turning to the concrete evaluation of individual contributions to the
expansion, one should estimate their  relevance, i.e. the powers in $D$
that are to be expected. To this end, consider a term of $n$th order in $b$,
distributed over $l$ traces of $n_j$th order, $n=\sum_{j=1}^l n_j$.  We then have $n$
sums over $\mu$ at the bare level. Each trace has its own $\mu$-conservation which
brings the number down to $n-l$. Now perform $n/2-1$ contractions over all but two
$b$'s. Each contraction removes one free summation, and we are down to
$\frac{n}{2}-l+1$ free summations. One of the sums is used for the un-contracted
index entering the quadratic action, which means that we are to expect a contribution
of order $D^{n-2l}$ (each sum over $\mu$ has $D^2$ terms, which is the dimension 
of the matrix Hilbert space $V \otimes V^*$). 
This power is reduced by the pre-factor $D^{-n/2}$
weighing $n$ expansions of fluctuation matrices as in
Eq.~\eqref{eq:Action_b_expanded}. Finally, the $l$ traces of the starting expression
contribute a factor $D^{l}$. We conclude that, pending other constraints, a maximum
power of $D^{\frac{n}{2}-l}$ is to be expected. This estimate indicates
that at $n$th order of perturbation theory, contributions of highest order come from terms where all $n$ fluctuation matrices enter the same trace, i.e.
from first order expansions of the exponentiated action in traces of maximal order. If the full power $D^{\frac{n}{2}-1}$ would result from the contraction of these traces, the perturbation theory would blow up. However, as we shall see, the  commutation relations between $X_\mu$-operators lead to a further reduction and bring the terms down to a small contribution of $\mathcal{O}(1)$.

Turning to the concrete evaluation of single traces, $\textrm{tr}(b_{\mu_1}\dots
\tilde b_{\mu_n})$, consider the example of  a 10th order trace
graphically represented in the bottom panel of  Fig.~\ref{fig:ContractionRules},
where the open circles represent the two uncontracted $b$-matrices. A first thing to
notice is that contractions with parallel contraction lines are vacuum contributions
and vanish. This is seen from $\langle \textrm{tr}(X b_\mu \tilde b_\nu Y b_\nu
\tilde b_\mu)\rangle\to \langle \textrm{tr}(Y)\textrm{tr}(Xb_\mu \tilde b_\mu)\rangle
\to \textrm{tr}(X)\textrm{tr}(Y)\textrm{tr}(1)\rangle$. We thus focus on the
`maximally crossed' structures shown in the right panel. (These should not be
confused with the maximally crossed diagrams of weak localization theory. In the
present context, each wavy line represents an `SYK-diffuson' and not a single
impurity line as in localization theory.)

It is straightforward to verify that the maximally crossed contraction of a trace of order $n=2(2k+1)$ generates the contribution
\begin{align}
     X_{k}\equiv D^{-(2k+1)} \sum_{\mu}\textrm{tr}(b_{\mu_1}\tilde b_{\mu_1}) \prod_{j=2}^{2k+1}s(\mu_j)\Pi_{\mu_j} \textrm{tr}_V(X_{\mu_1}X_{\mu_2}\dots X_{\mu_{2k+1}}X_{\mu_1}X_{\mu_2}\dots X_{\mu_{2k+1}}),
\end{align}
where $\sum_\mu\equiv \sum_{\mu_1,\dots,\mu_{2k+1}}$ is a sum over all index configurations.
Each operator $X_l$ appears twice under the trace and we commute them through the other $X$-operators to annihilate them as $X_l^2=s_l$. This leads to the appearance of a factor

\begin{align*}
    \textrm{tr}_V(X_{\mu_1}X_{\mu_2}\dots X_{\mu_{2k+1}}X_{\mu_1}X_{\mu_2}\dots X_{\mu_{2k+1}})=Ds(\mu_1) \prod_{j=2}^{2k+1} (s(\mu_{j},\nu_{j-1}) s(\mu_j)),\qquad \nu_j=\prod_{l=1}^{j}\mu_l,
\end{align*}
so that we obtain
\begin{align}
    \label{eq:XkBeforeMuSummation}
     X_{k}\equiv  D^{-2k}\sum_{\mu}s(\mu_1)\,\textrm{tr}(b_{\mu_1}\tilde b_{\mu_1}) \prod_{j=2}^{2k+1}(\Pi_{\mu_j}  s(\mu_j,\nu_{j-1})).
\end{align}
At this point, and excluding $\mu_1$, we are summing over $D^{4k}$ terms and may expect a contribution of total order $D^{-2k}\cdot D^{4k}=D^{\frac{n}{2}-1}$ in accordance with the estimate above. However, this estimate turns out to be way too high as it ignores an almost complete sign cancellation due to the presence of the propagators weighing the sum. 

In~\ref{sec:SignProduct} we show that the summation over $\mu$ can be carried out in closed form and that it  converts the sum into the sign factor
\begin{align}
    \label{eq:XkAfterMuSummation}
    X_{k}\equiv  \sum_{\mu}s(\mu_1)\,\textrm{tr}(b_{\mu_1}\tilde b_{\mu_1}) \left(\frac{1}{n^{2k}} \sum_{a_1,\dots,a_{2k}} \prod_{j=0}^{2k-2} s\left(\prod_{l=2k-j}^{2k} a_l,a_{2k-j-1}\right) \right)s\left(\prod_{j=1}^{2k} a_j,\mu_1\right).
\end{align}
The key feature of this expression is that (i) the summation over $2k$ indices $\mu$
did not lead to a contribution of $\mathcal{O}((D^{2})^{2k})$ but only
$\mathcal{O}(D^{2k})$. In combination with the factor $D^{-2k}$ up front, this makes
for a $D$-independent scaling, as for the `bare' term of quadratic order. However,
(ii) the summation also leads to a sum over $2k$ quartic configurations
$\sum_aF(a)\equiv \sum_{i,j,k,l}F(\chi_i\chi_j\chi_k\chi_l)$. Each sum is over
$n=\binom{N}{4}$ terms, and the total sum extends over a complicated, and effectively
random sign factor. This means that the typical value of the sum will be
$\mathcal{O}(\sqrt{n^{-2k}})=\mathcal{O}(n^{-k})=\mathcal{O}(e^{-4k\ln(N)})$. We therefore conclude that higher order terms in the perturbation expansion suffer exponential suppression and can be neglected. 

\section{Summary and discussion} % (fold)
\label{sec:summary_and_discussion}
In this paper we explored the approach to quantum ergodicity in the long time
dynamics of the SYK model. It turned out that the  irreversible relaxation to the
ergodic limit is governed by a large set of collective modes, each labeled by an
element of the $2^N/4$ dimensional restriction of the Clifford algebra to the even
parity sector.  We reasoned that these modes play a role conceptually analogous to
diffusion modes in a dirty single particle medium. The main differences were (a) an
density of states growing exponentially in the norm of the mode index, i.e. the
number of Majorana-generators contained in it, and (b) a relaxation rate quickly
increasing in the same index. The competition of these tendencies led to distinct
signatures in spectral correlations. Specifically, we observed that the large number
of modes generates a largely structureless contribution to the spectral two-point
function which masks the ergodic RMT profile already a low energies exceeding the
many body level spacing only by factors polynomial (and not exponential) in the state
number, $N$. A different perspective could be obtained from the time dependent
spectral form factor which filtered out the contribution of the modes of highest
longevity, and showed a strong enhancement over the RMT form factor at short times.
Both the results obtained for the spectral two point function and for the form factor
agree favorably with previous numerical work in a comparison that does not involve
adjustable parameters. In particular, the characteristic `dip-ramp' structure
observed in the form factor resembles results previously obtained for the OTO
correlation functions of the system (if only at $\beta\to 0$, the only limit
accessible to our analysis.)

Finally, one may wonder how the collective modes discussed here compare to the
conformal Goldstone mode fluctuations addressed in other works. It can be reasoned
that these two types of excitations focus on different sectors of the system's phase
space. For example, the effective Liouville quantum mechanics~\cite{Bagrets:2016} describing the
conformal symmetry breaking in the system ignores replica off-diagonal fluctuations
(fluctuations off-diagonal in replica space and/or replica symmetry breaking do not
play a role), while they are essential in the present context. As a direct
consequence, the RMT `propagator', $\omega^{-1}$, governing the zero mode action in
the present theory is nowhere in sight in the Liouville theory. In other words, the
latter cannot describe the ergodic limit of quantum chaos. Conversely, the
time-dependent conformal symmetry is not included in the present framework, which
singles out two fixed frequencies from the beginning. Still it would be desirable to
understand the connection between the two classes of fluctuations at a deeper level
and this is a subject of further work.

\vskip.5cm
\noindent  \emph{Acknowledgments:} We thank Alex Kamenev for discussions and Mazaki Tezuka for sharing information on the numerical aspects of Ref.~\cite{Cotler:2017}. Work supported by CRC 183 of the Deutsche Forschungsgemeinschaft (project A03).

%%%%%%%%%%%%%%%%%%%%%%%%%%%%%%%%%%%%%%%%%%%%%%%%%%%%%%%%%%%%%%%%%%%%%%%%%%%%%%%%%%%
\appendix

\section{Derivation of Eq.~\eqref{eq:ActionInhomogeneousModesIntegrated}} 
\label{sec:DerInhomogeneousmodesIntegrated}
In this Appendix we detail the integration over the inhomogeneous modes $a_{a,\mu}$
leading from Eq.~\eqref{eq:ZBeforeIntInh}
to~\eqref{eq:ActionInhomogeneousModesIntegrated}. A  rearrangement of terms brings
the Gaussian  action of the functional integral into the form,
\begin{align*}
    S_\mathrm{g}[a]=- \frac{1}{2n} \sum_{a,\mu}  s(\mu) s(a,\mu) \mathrm{tr}\left (\bar a_\mu^2+ 2(\lambda_{\mu}s(a,\mu)+\bar a_\mu)a_{a,\mu}+a_{a,\mu}^2\right).
\end{align*}
The quadratic integration over $a_{a,\mu}$ then leads to
\begin{align*}
    S_\mathrm{g}[a]&= -\frac{1}{2n} \sum_{a,\mu}  s(\mu) s(a,\mu) \mathrm{tr}\left (\bar a_\mu^2- (\lambda_{\mu}s(a,\mu)+\bar a_\mu)^2\right)=\cr
    &=\frac{1}{2} \sum_{\mu}  s(\mu) \mathrm{tr}\left( S(\mu)\lambda_\mu^2 +2  \lambda_\mu \bar a_\mu\right),
\end{align*}
where $s(a,\mu)^2=1$ was used. We finally integrate over $\lambda_\mu$ to obtain
$S_\mathrm{g}[a]=-\frac{1}{2}\sum_{\mu}  s(\mu) S(\mu)^{-1} \mathrm{tr}(\bar a_\mu^2)$.
Dropping the bar, we arrive at the Gaussian contribution to
Eq.~\eqref{eq:ActionInhomogeneousModesIntegrated}.

\section{Derivation of the RMT spectral correlation function} % (fold)
\label{sec:RMTSpectralCorrelation}

Referring to Ref.~\cite{Kamenev:1999} for details we here sketch how the RMT spectral
correlation function Eq.~\eqref{eq:R2RMT} is obtained from the integration of the
effective action~\eqref{eq:SSigmaModel} over $Q$. Referring to Ref.~\cite{Zirnbauer:1996} we first
note that the convergence of the integral requires a restriction of the integration
domain to the subset  $\mathrm{U}(2R,2R)/\mathrm{U}(R,R)\times\mathrm{U}(R,R)\subset
\mathrm{GL}(2R,2R)/\mathrm{GL}(R,R)\times\mathrm{GL}(R,R)$ of the full coset symmetry
manifold. Second, the integral $\int dQ\exp(-S[Q])$ over the left-invariant measure
is `semiclassically exact' (in the sense of equivariant cohomology~\cite{Szabo:2000}). In practical
terms, this means that it can be rigorously computed by stationary phase methods.
Here, the terminology stationary phase refers to the extremal configurations
$\delta_Q S[Q]=0$ on the Goldstone mode manifold, not to be confused with the
stationary phase approximation which introduced the $Q$-degrees of freedom in the
first place. In a magnetic analogy, the fluctuations of $Q$ play the role of
magnon-fluctuations, $\hat z$ is analogous to a weak explicitly symmetry breaking
fields, and the  stationary points of $Q$ correspond to magnetization axes aligned
with that field.

To identify the stationary configurations we note that fluctuations around any configuration $Q=T \tau_3 T^{-1}$ are generated by $Q\to (T \delta T) \tau_3 (\delta T^{-1}T)$, where 
\begin{align}
    T=\exp(W)\equiv \exp \left( \begin{matrix}
        &B\cr -B^\dagger &
    \end{matrix} \right), 
\end{align}
and the off-diagonal matrix form in advanced-retarded space implements the coset structure (much like in a magnet fluctuations around the $z$-axis configuration $\sigma_3$ have magnon generators $\sigma_{x,y}$.) and the anti-hermitean generators $W$ are parameterized by complex replica space matrices $B\in \textrm{GL}(R,R)$.  We verify that the variation of the action $\delta_BS[Q]=0$ leads to the equation $[Q,\hat z]=0$, which is equivalent to $[Q,\tau_3]=0$. This equation has the natural (causal) solution $Q=\tau_3$, which in the jargon of the field is called the standard saddle point. A quadratic expansion of the action action around this configuration leads to the Gaussian action
\begin{align}
    S[B,B^\dagger]=-i\frac{\pi \Delta z}{\Delta}\textrm{tr}(BB^\dagger),
\end{align}
where $\Delta z=z^+-z^- =\omega^+ +(h_+-h_-)$. Note that the convergence of the $B$-integral is  safeguarded by the imaginary increment in $\omega$. The Gaussian fluctuation over the $R^2$ matrix elements of $B$ then leads to the result
\begin{align}
    Z_0(h)= \left(-i\frac{\pi (\omega^++h_+-h_-)}{\Delta}\right)^{-R^2},
\end{align}
where we noted that the saddle point action $S[\tau_3]\propto R$ vanishes in the replica limit. 

From here, the correlation function $C(\omega)$ is obtained by differentiation w.r.t. $h_\pm$. Taking the replica limit, we note that the only surviving contribution reads
\begin{align}
    C_0(\omega)=-\left( \frac{1}{ \omega^+} \right)^2, 
\end{align}
and from here the spectral two point function, $R_2(\omega)=\frac{\Delta^2}{2\pi^2}  \textrm{Re}(C_0(\omega))$ is obtained as $R_{2}(\omega)=-\frac{\Delta^2} {2\pi^2} \textrm{Re}\frac{1}{\omega^{+2}}$.
This is identical to the large energy, $\omega>\Delta$ approximation to the RMT spectral correlation function and  to  the $k=0$ contribution to the correlation function Eq.~\eqref{eq:R2Modes}. 

For completeness, we quickly outline how the full non-perturbative (in the parameter
$(\omega/\Delta)^{-1}$) RMT correlation function is obtained from the theory~\cite{Kamenev:1999,Andreev:1995}.
Contributions beyond $C_0$ emerge from non-causal solutions to the stationary phase
equation $[Q,\tau_3]=0$. The equation is solved by any configuration $\bar Q=\tau_3 \otimes
S$, where $S^{rr'}=\delta^{rr'}s_r$ an arbitrary matrix of sign factors $s_r=\pm 1$.
However, a straightforward computation of the corresponding fluctuation determinants
shows that the majority of these configurations leads to fluctuation determinants vanishing in the replica limit. The only survivor contributions are matrices containing a sign flip in only one replica channel, say $\bar Q\tau_3\otimes (-2P^1+1)$, where $P^1$ projects on the first replica. The source-free (the sources are needed in the fluctuation determinants around the saddle point to obtain contributions non-vanishing in the replica limit)  action of these configurations, $S[\bar Q]=\frac{i\pi}{2\Delta}\textrm{tr}(\bar Q \tau_3\hat \omega)=\frac{i \pi \omega^+}{2\Delta}\mathrm{tr}(-2 P^1+1)= \frac{i \pi \omega^+}{2\Delta}\mathrm{tr}(-2 P^1+1)=\frac{i \pi \omega^+}{\Delta}(-2+R)\stackrel{R\to 0}\longrightarrow - \frac{i2 \pi \omega^+}{\Delta}$ no longer vanishes in the replica limit. 

A straightforward analysis of fluctuations around these points %(see Ref.~\cite{Kamenev:1999}  for details) 
shows that the fluctuation determinant remains the same, up to a global minus sign. The two point function is thus obtained as
\begin{align}
    R_{2}(\omega)=-\frac{1}{2}\left(\frac{\Delta}{\pi}\right)^2\textrm{Re}\frac{1}{\omega^{+2}}\left( 1-e^{\frac{i2 \pi \omega^+}{\Delta}} \right)=\Delta \delta(\omega)-\left( \frac{\sin(\pi \omega^+/\Delta)}{\pi\omega^+/\Delta} \right)^2, 
\end{align}
which is the RMT result. 

\section{From Eq.~\eqref{eq:GaussianActionMassive} to Eq.~\eqref{eq:GaussianActionMassiveEffective}} 
\label{sec:GaussianActionMassive}
We here discuss the derivation of Eq.~\eqref{eq:GaussianActionMassiveEffective} is derived from the precursor Eq.~\eqref{eq:GaussianActionMassive}.  The anti-commutativity of the off-diagonal fluctuations, $w_\mu$ with $\tau_3$  implies that
\begin{align*}
    S_\textrm{m}[w]=-\frac{1}{2}\sum_\mu s(\mu) \,\mathrm{tr}\left(\left[ S(\mu)^{-1}- \left( \frac{\hat z}{2\gamma}-i \tau_3 \sqrt{1- \left( \frac{\hat z}{2\gamma} \right)^2  } \right)   \left( \frac{\hat z^\ast}{2\gamma}+i \tau_3 \sqrt{1- \left( \frac{\hat z^\ast}{2\gamma} \right)^2  } \right)   \right]w_\mu^2\right).
\end{align*}
where we defined $\hat z^\ast \equiv \tau_1 \hat z \tau_1 = \epsilon -
(\frac{\omega}{2}+i\delta)\tau_3 +\textrm{diag}(h_-,h_+)$, i.e. the matrix $\hat z$
with diagonal matrix elements exchanged. Compared to the scales, $\epsilon,\gamma$
the arguments, $\omega,h_\pm$ contained in $\hat z$ are weak, and so it makes sense
to expand the $\hat z$-dependent factors to first order in these quantities. The expansion of the  product containing the $\sqrt{}$-factors yields $\left(\dots\right)
\left( \dots \right)\simeq 1 +\frac{i}{\gamma}\big( 1-
\big(\frac{\epsilon}{2\gamma}\big)^2  \big)^{-1/2} (\omega+(h_+-h_-)) 
 $, and we obtain
\begin{align*}
    S_\textrm{m}[w]\simeq - \frac{1}{2}\sum_\mu s(\mu) \,\mathrm{tr}\left(\left[
    S(\mu)^{-1}-1-i\frac{\omega+(h_+-h_-)}{\gamma\sqrt{1- \left(
    \frac{\epsilon}{2\gamma} \right)^2  }}  \right]w_\mu^2\right).
\end{align*}
Finally, using that close to the band center $\big(1-
\big(\frac{\epsilon}{2\gamma}
\big)^2  \big)^{-1/2}\simeq 1$ we obtain Eq.~\eqref{eq:GaussianActionMassiveEffective}.

\section{Derivation of Eq.~\eqref{eq:XkAfterMuSummation}} % (fold)
\label{sec:SignProduct}
In this Appendix we show how do do the $\mu$-summation in Eq.~\eqref{eq:XkBeforeMuSummation} to obtain Eq.~\eqref{eq:XkAfterMuSummation}. 
The  key relation required in the process reads
 \begin{align}
     \sum_\mu s(\mu,\nu)=D^2\delta_{\nu,0}.
 \end{align}
It states that for every non-trivial operator $\nu$ there are equally many commuting and anti-commuting operators in the algebra. The sum weighed by the commutation signs thus vanishes, unless $\nu=1$ is the identity, in which case it yields the dimension $D^2$ of the algebra. 

Let us now consider the prefactor of the quadratic action in Eq.~\eqref{eq:XkBeforeMuSummation}, 
\begin{align}
    \label{eq:PkStarting}
    P_{k}\equiv  D^{-2k}\sum_{\mu_2\dots\mu_{2k+1}} \prod_{j=2}^{2k+1}\Pi(\mu_j)  s\left(\mu_j,\nu_{j-1}\right).
\end{align}
 For simplicity, we assume that the propagator is dominated by the zeroth order expansion in  $1<|S(\mu)|^{-1}$, $\Pi(\mu)=(S(\mu)^{-1}-1)^{-1}\simeq S(\mu)$. One can convince oneself that this is a conservative estimate and that higher order terms lower the result. 

Below we will show that if $2l$ out of $2k+1$ summations over $\mu$-indices are performed, the sum assumes the form
\begin{align}
    \label{eq:Pkl}
    P_k&=D^{-2(k-l)}\sum_{\mu_2\dots \mu_{2k+1-2l}}X_{2l}(\mu)\frac{1}{n^{2l}}\sum_{a_{2k-2l+1}\dots a_{2k}} Y_{2l}(a)s(b_{2k-2l+1},\nu_{2k+1-2l}),\cr
    &\quad X_{2l}(\mu)=\prod_{j=2}^{2k+1-2l}S(\mu_j)s(\mu_j,\nu_{j-1}),\cr
    &\quad Y_{2l}(a)=\prod_{j=0}^{2l-2}s(b_{2k-j},a_{2k-j-1}),
\end{align}
where
\begin{align}
    \label{eq:NuAndBdef}
    \nu_j=\prod_{i=1}^j\mu_l,\qquad b_j=\prod_{i=j}^{2k} a_i.
\end{align}
 For $l=k$, no
$\mu$-summation is left, $X_{2k}=1$, and the remaining terms reduce to
Eq.~\eqref{eq:XkAfterMuSummation}. The formula above is proven by induction. For
$l=0$, we have the starting expression~\eqref{eq:PkStarting}. Let us then assume that
the expression holds for a value $2l$, and do two more $\mu$-summations to progress
to $2l+2$. We first sum over $\mu_{2k+1-2l}\equiv \tilde \mu$. The dependence on this
index sits in the highest factor $S(\tilde \mu)s(\tilde \mu,\nu_{2k-2l})$
contributing to $X_{2l}$ and in the sign factor
$s(b_{2k-2l+1},\nu_{2k+1-2l})=s(b_{2k-2l+1},\tilde
\mu\nu_{2k-2l})=s(b_{2k-2l+1},\nu_{2k-2l})s(b_{2k-2l+1},\tilde \mu)$. Isolating these
terms and using $S(\tilde \mu)=\frac{1}{n}\sum_{a_{2k-2l}}s(a_{2k-2l},\tilde \mu)$,
we obtain
\begin{align}
    P_k&=D^{-2(k-l)}\sum_{\mu_2\dots \mu_{2k-2l}}X_{2l+1}(\mu)\frac{1}{n^{2l+1}}\sum_{a_{2k-2l}\dots a_{2k}} Y_{2l}(a)s(b_{2k-2l+1},\nu_{2k-2l})\times\cr
    &\qquad\times \sum_{\tilde \mu}s(b_{2k-2l+1},\tilde \mu)s(a_{2k-2l},\tilde \mu)s(\nu_{2k-2l},\tilde \mu).
\end{align}
we now note,
\begin{align}
   & \sum_{\tilde \mu}s(b_{2k-2l+1},\tilde \mu)s(a_{2k-2l},\tilde \mu)s(\nu_{2k-2l},\tilde \mu)=\sum_{\tilde \mu}s(b_{2k-2l+1}a_{2k-2l}\nu_{2k-2l},\tilde \mu)=\cr
    &\qquad =D^2 \delta_{b_{2k-2l+1}a_{2k-2l}\nu_{2k-2l},0}=D^2\delta_{b_{2k-2l}\nu_{2k-2l},0}=D^2\delta_{\mu_{2k-2l},b_{2k-2l}\nu_{2k-2l-1}}
\end{align}
to conclude that the sum over sign factors collapses one more $\mu$-sum, i.e. the sum over $\mu_{2k-2l}$. Isolating the dependence of the summand on this parameter, and using the $\delta$-constraint, we obtain
\begin{align}
   P_k&=D^{-2(k-(l+1))}\sum_{\mu_2\dots \mu_{2k+1-2(l+1)}}X_{2(l+1)}(\mu)\frac{1}{n^{2(l+1)}}\sum_{a_{2k+1-2(l+1)}\dots a_{2k}} Y_{2l}(a)\times\cr
&\times   \left[s(b_{2k-2l+1},\mu_{2k-2l}\nu_{2k-2l-1})s(a_{2k-2l-1},\mu_{2k-2l})s(\nu_{2k-2l-1},\mu_{2k-2l})\right]_{\mu_{2k-2l}=b_{2k-2l}\nu_{2k-2l-1}}.
\end{align}
Using the constraint, the term is angular brackets becomes
\begin{align}
    s(b_{2k-2l+1},b_{2k-2l})s(a_{2k-2l-1},b_{2k-2l}\nu_{2k-2l-1})s(\nu_{2k-2l-1},b_{2k-2l}\nu_{2k-2l-1}).
\end{align}
We manipulate the factors appearing in this product as
\begin{align}
    &s(b_{2k-2l+1},b_{2k-2l})=s(b_{2k-2l+1},a_{2k-2l}b_{2k-2l+1})=s(a_{2k-2l},b_{2k-2l+1}),\cr
    &s(a_{2k-2l-1},b_{2k-2l}\nu_{2k-2l-1})=s(a_{2k-2l-1},b_{2k-2l})s(a_{2k-2l-1},\nu_{2k-2l-1}),\cr
    &s(\nu_{2k-2l-1},b_{2k-2l}\nu_{2k-2l-1})=s(\nu_{2k-2l-1},b_{2k-2l}).
\end{align}
Including these factors into the terms of the sum we obtain
\begin{align}
   P_k&=D^{-2(k-(l+1))}\sum_{\mu_2\dots \mu_{2k+1-2(l+1)}}X_{2(l+1)}(\mu)\frac{1}{n^{2(l+1)}}\sum_{a_{2k+1-2(l+1)}\dots a_{2k}} Y_{2(l+1)}(a)s(b_{2k+1-2(l+1)},\nu_{2k+1-2(l+1)}),
\end{align}
which is the original expression with a replacement $l\to l+1$. If we now set $k=l$ in Eq.~\eqref{eq:Pkl}, we obtain
\begin{align}
    P_k=\frac{1}{n^{2k}}\sum_{a_1\dots a_{2k}}\prod_{j=0}^{2k-2}s(b_{2k-j},a_{2k-j-1})s(b_1,\mu_1),
\end{align}
where $\nu_1=\mu_1$ was used. Recalling the definition~\eqref{eq:NuAndBdef}, we obtain Eq.~\eqref{eq:XkAfterMuSummation}.

% section derivation_of_eq_ (end)

% section from_eq_eq:gaussianactionmassive (end)

%\vspace{0.5cm}
%\noindent{\bf References}\\
%%\bibliographystyle{apsrev-nourl}
%%\bibliographystyle{plain} 
\bibliographystyle{unsrt}

%\bibliography{../../../revtex/BIBLIOGRAPHY/library}{}
\bibliography{Bibliography}
%\bibliography{library}


%\begin{thebibliography}{00}



%\end{thebibliography}
\end{document}